\documentclass[usenatbib]{mnras}
\usepackage{textcomp}
\usepackage[utf8]{inputenc}
\usepackage{lscape}
\usepackage{amsmath,amssymb}
\usepackage{mathrsfs}	
\usepackage{multicol}
\usepackage[dvipdfmx]{graphicx}
\usepackage{caption}
\usepackage{subcaption}
\graphicspath{{figures/}}

\usepackage{times}
\usepackage{rotating}
\usepackage{ulem}
\catcode`\@=11
\def\gta{\ifmmode{\,\mathrel{\mathpalette\@versim>\,}}
    \else{$\,\mathrel{\mathpalette\@versim>}\,$}\fi}
\def\lta{\ifmmode{\,\mathrel{\mathpalette\@versim<\,}}
    \else{$\,\mathrel{\mathpalette\@versim<}\,$}\fi}
\def\@versim#1#2{\lower 2.9truept \vbox{\baselineskip 0pt \lineskip
    0.5truept \ialign{$\m@th#1\hfil##\hfil$\crcr#2\crcr\sim\crcr}}}
\catcode`\@=12

{\newif\ifnotend
\notendtrue
\def\veclist{ABCDEFGHIJKLMNOPQRSTUVWXYZabcdefghijklmnopqrstuvwxyz.}
\def\top#1#2.{#1}
\def\tail#1#2.{#2.}
\loop\expandafter\xdef\csname bb\expandafter\top\veclist\endcsname%
{{\noexpand\bf\expandafter\top\veclist}}
\edef\veclist{\expandafter\tail\veclist}
\if\veclist.\notendfalse\fi\ifnotend\repeat}

\defcitealias{Rownd1994}{R04}
\defcitealias{Kornreich2000}{K00}
\defcitealias{Epinat2008}{E08}
\defcitealias{Bellazzini2020}{B20}

\newcommand{\ldm}	{l_{\rm dm}}
\newcommand{\lstar} {l_{\star}}
\newcommand{\lgas}  {l_{\rm gas}}
\newcommand{\lPB}   {l_{\star, {\rm PB}}}
\newcommand{\lPBdm} {l_{{\rm dm, PB}}}
\newcommand{\Nh}	{N_{\rm halo}}
\newcommand{\Ngas}	{N_{\rm gas}}
\newcommand{\Nstar} {N_{\star}}
\newcommand{\Nb}	{N_{\rm PB}}
\newcommand{\Nbdm}	{N_{\rm dm, PB}}
\newcommand{\MdmPB}	{M_{\rm dm, PB}}

\newcommand{\tmax}	{t_{\rm max}}
\newcommand{\tint}  {t_{\rm int}}
\newcommand{\tc}    {t_{\rm cross}}
\newcommand{\tdf}   {t_{\rm fric}}
\newcommand{\rt}    {r_{\rm t}}
\newcommand{\Smax}  {\Sigma_{\rm max}}
\newcommand{\Phieff}    {\Phi_{\rm eff}}
\newcommand{\bbRPB}     {{\bf R_{\rm PB}}}
\newcommand{\Rsnapi}    {R_{{\rm s}, i}}
\newcommand{\Sigsnapi}  {\Sigma_{{\rm s}, i}}
\newcommand{\mpart}     {m_{\rm part}}
\newcommand{\xii}    {x_i}
\newcommand{\yi}    {y_i}
\newcommand{\zi}    {z_i}

\newcommand{\Lz}	{L_z}

\newcommand{\Ri}	{R_i}
\newcommand{\Rk}	{R_k}
\newcommand{\nnk}	{n_k}

\newcommand{\vk}	{v_k}

\newcommand{\sigmak}{\sigma_k}
\newcommand{\HI}	{{\rm H\textsc{i}}}
\newcommand{\HII}   {{\rm H\textsc{ii}}}

\newcommand{\Halpha}{{\rm H}\alpha}
\newcommand{\Ns}	{N_n}
\newcommand{\Nv}	{N_v}
\newcommand{\tth}	{{\rm th}}

\newcommand{\mprot}	{m_{\rm p}}
\newcommand{\nint}	{n_{\rm int}}
\newcommand{\nobs}	{n_{\rm obs}}

\newcommand{\UpsilonPB}     {\Upsilon_{\rm PB}}
\newcommand{\Upsilondisc}   {\Upsilon_{\rm disc}}

\newcommand{\bbxi}	{{\boldsymbol \xi}}

\newcommand{\Mgas}	    {M_{\rm gas}}
\newcommand{\hgas}	    {h_{\rm gas}}
\newcommand{\Mstar}	    {M_{\star}}
\newcommand{\rhostar}   {\rho_{\star}}
\newcommand{\hstar}	    {h_{\star}}
\newcommand{\zstar}     {z_{\star}}
\newcommand{\vgas}	    {v_{\rm gas}}
\newcommand{\vstar}	    {v_{\star}}
\newcommand{\Many}      {M_{\rm cmp}}
\newcommand{\hany}      {h_{\rm cmp}}
\newcommand{\yany}      {y_{\rm cmp}}
\newcommand{\vany}      {v_{\rm cmp}}
\newcommand{\Sigmastar} {\Sigma_{\star}}
\newcommand{\Sigmagas}  {\Sigma_{\rm gas}}
\newcommand{\rhogas}    {\rho_{\rm gas}}

\newcommand{\hvRii}     {\langle v_R^2 \rangle }
\newcommand{\hvzii}     {\langle v_z^2 \rangle }
\newcommand{\hvphi}     {\langle v_{\phi} \rangle }
\newcommand{\hvphiii}   {\langle v_{\phi}^2 \rangle}

\newcommand{\vphigas}   {v_{\phi,\rm gas}}

\newcommand{\Magn}      {{\rm M}}

\newcommand{\rhos}	{\rho_{\rm s}}
\newcommand{\rs}	{r_{\rm s}}
\newcommand{\Mvir}	{M_{\rm vir}}
\newcommand{\rvir}	{r_{\rm vir}}
\newcommand{\Mtot}	{M_{\rm tot}}
\newcommand{\rhodm} {\rho_{\rm dm}}
\newcommand{\Mdm}	{M_{\rm dm}}
\newcommand{\Phidm} {\Phi_{\rm dm}}
\newcommand{\vh}	{v_h}

\newcommand{\Phitot}{\Phi_{\rm tot}}
\newcommand{\vc}	{v_{\rm c}}

\newcommand{\vcmax}	{v_{\rm c, max}}

\newcommand{\PhiPB}     {\Phi_{\rm PB}}
\newcommand{\PhistarPB} {\Phi_{\star, \rm PB}}
\newcommand{\PhidmPB}   {\Phi_{\rm dm, PB}}
\newcommand{\PhiastPB}  {\Phi_{*, {\rm PB}}}
\newcommand{\rhostarPB} {\rho_{\star, \rm PB}}
\newcommand{\rhodmPB}   {\rho_{\rm dm, PB}}
\newcommand{\rhoastPB}  {\rho_{*, {\rm PB}}}
\newcommand{\fstarPB}   {f_{\star, {\rm PB}}}
\newcommand{\fdmPB}     {f_{\rm dm, PB}}
\newcommand{\fastPB}    {f_{*, {\rm PB}}}
\newcommand{\Mb}	    {M_{\rm PB}}
\newcommand{\Ie}	    {I_{\rm e}}
\newcommand{\Ltot}  	{L_{\rm tot}}
\newcommand{\Reff}	    {R_{\rm e}}
\newcommand{\bm}	    {b_m}
\newcommand{\rh}	    {r_{\rm h}}
\newcommand{\rsPB}      {r_{\rm s, PB}}
\newcommand{\rtPB}      {r_{\rm t, PB}}
\newcommand{\rhosPB}    {\rho_{\rm s, PB}}
\newcommand{\MvirPB}    {M_{\rm vir, PB}}
\newcommand{\MdynPB}    {M_{\rm dyn, PB}}

\newcommand{\sigmar}	{\sigma_r}
\newcommand{\sigmaphi}	{\sigma_\phi}
\newcommand{\sigmatheta}{\sigma_\theta}

\newcommand{\sigmai}    {\sigma_i}

\newcommand{\RPB}       {R_{\rm PB}}
\newcommand{\Sersic}    {S\'{e}rsic\,}
\newcommand{\Imax}      {I_{\rm max}}

\newcommand{\kpc}	{\, {\rm kpc}}
\newcommand{\Mpc}	{\, {\rm Mpc}}
\newcommand{\Myr}   {\, {\rm Myr}}
\newcommand{\Gyr}	{\, {\rm Gyr}}
\newcommand{\K}     {\, {\rm K}}
\newcommand{\Msun}	{\, M_{\odot}}
\newcommand{\kms}	{\, {\rm km \,s^{-1}}}
\newcommand{\asec}	{\, {\rm arcsec}}
\newcommand{\amin}  {\, {\rm arcmin}}

\newcommand{\sech}  {{\rm sech}}
\newcommand{\LL}	{\mathcal{L}}
\newcommand{\data}	{\mathcal{D}}
\newcommand{\dd}	{\text{d}}
\newcommand{\DD}	{\partial}
\newcommand{\logten}{\log_{10}}
\newcommand{\rapex}	{\textquoteright\,}
\newcommand{\lapex}	{\textquoteleft}

\newcommand{\comp}  {{\rm cmp}}
\newcommand{\dm}    {{\rm dm}}
\newcommand{\disc}  {{\rm disc}}
\newcommand{\gas}   {{\rm gas}}

\newcommand{\AREPO}	{{\rm \textsc{arepo}}\,}

\begin{document}

\date{Accepted 2020 November 24. Received 2020 November 24; in original form 2020 August 28}

\pubyear{2020}

\title[Hydrodynamical $N$-body models of NGC 5474]{An off-centred bulge or a satellite? \\ Hydrodynamical $N$-body simulations of the disc galaxy NGC 5474}
{}
\author[R. Pascale et al.]{R. Pascale$^1$\thanks{E-mail: raffaele.pascale@inaf.it}, M. Bellazzini$^1$, M. Tosi$^1$, F. Annibali$^1$, F. Marinacci$^2$, C. Nipoti$^{1,2}$
\\ \\
$^1$INAF - Osservatorio di Astrofisica e Scienza dello Spazio di Bologna, Via Gobetti 93/3, 40129 Bologna, Italy \\ 
$^2$Dipartimento di Fisica e Astronomia, Università di Bologna, via Gobetti 93/2, 40129, Bologna, Italy\\}

\label{firstpage}
\pagerange{\pageref{firstpage}--\pageref{lastpage}}

\maketitle

\begin{abstract}
We present dynamical models of the star-forming galaxy NGC 5474 based on $N$-body hydrodynamical numerical simulations. We investigate the possible origin of the compact round stellar structure, generally interpreted as the bulge of the galaxy, but unusually off-set by $\simeq1\kpc$ in projection from the visual and the kinematic centres of both the star and the gas discs. We argue that it is very unlikely that the putative bulge is in a co-planar orbit in the disc plane, showing that such a configuration would be hardly compatible with its smooth and regular spatial distribution, and, in case its mass is above $10^8\Msun$, also with the regular $\HI$ velocity field of NGC 5474. Instead, if the putative bulge is in fact an early-type satellite galaxy orbiting around NGC 5474, not only the off-set can be easily produced by projection effects, but our simulations suggest that the gravitational interaction between the two systems can explain also the warped $\HI$ distribution of NGC 5474 and the formation of its loose spiral arms. As a by-product of the simulations, we find that the peculiar over-density of old stars detected in the south-west region of NGC 5474 may be explained with the interaction between NGC 5474 and a smaller stellar system, unrelated to the putative bulge, accreted in the disc plane.

\end{abstract}


\section{Introduction}
\label{sec:int}

Located at a distance of $6.98\Mpc$ \citep{Tully2013}, NGC 5474 is a local star forming galaxy, classified as SAcd pec, belonging to the M 101 Group. With an absolute blue magnitude $\Magn_B\simeq-18.4$, it is among the most luminous satellites of M 101, also known as the Pinwheel Galaxy, and it is also relatively close to it, with an angular separation smaller than $1^{\circ}$ \citep[corresponding to $\sim120\kpc$, in projection;][]{Tikhonov2015}.

Due to its various asymmetries \citep{Kornreich1998}, NGC 5474 was early recognized as peculiar \citep{Huchtmeier1979}. The $\HI$ disc is distorted at radial distances $R$ larger than $5\kpc$, resulting in a change of the position angle (PA) of $\simeq50^{\circ}$ twisting from $155^{\circ}$ for $R<5\kpc$ to $105^{\circ}$ for $R>8.5\kpc$ \citep[hereafter R04]{Rownd1994}. The change in the PA is thought to be associated with a warp in the $\HI$ disc, which connects the gaseous component to the south-western edge of M 101. This $\HI$ bridge \citep{Huchtmeier1979,vanderHulst1979} has been often considered as a tidal debris formed during a recent fly-by of NGC 5474 close to M 101 \citep{Mihos2012}. 

\begin{figure}
    \centering
    \includegraphics[width=1\hsize]{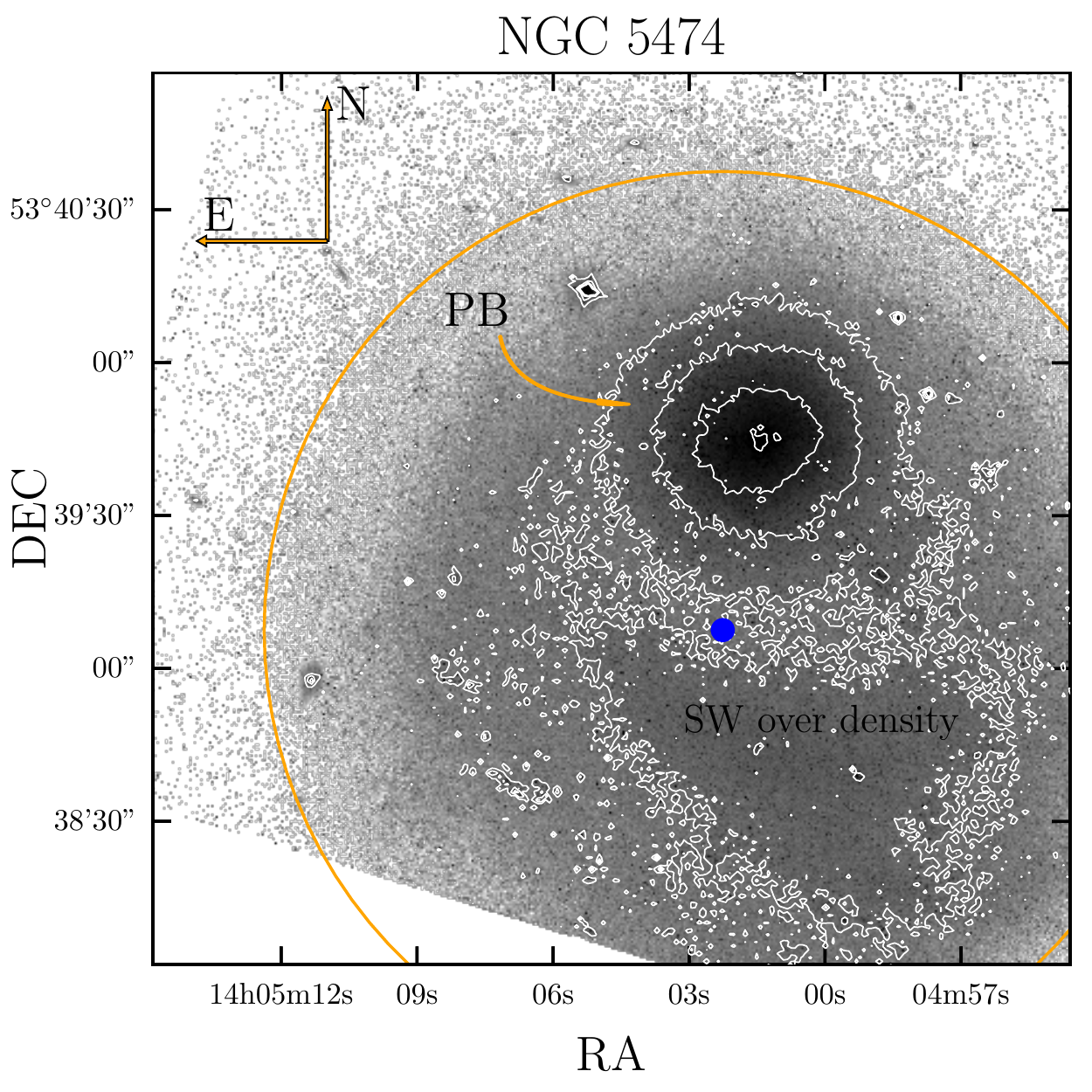}
    \caption{Image of NGC 5474 obtained using  observations in the F814W band from the LEGUS photometric catalogue (\citealt{Calzetti2015}). The blue dot shows the position of the $\HI$ kinematic center from \citetalias{Rownd1994}, the orange circle is centred on the $\HI$ kinematic center and its radius is $90\asec$ long, corresponding to $3\kpc$ at a distance $D=6.98\Mpc$. The white contours are separated by $10^{k/3-1}\Imax$, where $\Imax$ is approximately the surface brightness of the PB centre, while $k=0,1,2,3$. The orange arrow points the PB, while the SW over-density is to the South.  North is up and East is to the left.}\label{fig:ngc5474}
\end{figure}

The $1\kpc$ off-set between the kinematic centre of its $\HI$ disc and the optical centre of what has always been interpreted as the galaxy bulge is another fascinating peculiarity of NGC 5474. As an example, Fig.~\ref{fig:ngc5474} shows a zoomed-in view of the central region of NGC 5474 in the F814W band obtained from the LEGUS photometric catalogue \citep{Calzetti2015}: the kinematic centre is marked with a blue dot, while the off-set bulge is clearly visible to the north of the kinematic centre. First observed by \cite{vanderHulst1979}, then validated by \citetalias{Rownd1994} and \citet[hereafter K00]{Kornreich2000} looking at the $\HI$ emission, the off-set has also been confirmed from the $\Halpha$ kinematics \citep[hereafter E08]{Epinat2008}, making the picture even more puzzling. This oddly large discrepancy has recursively raised the question about the true nature of such a stellar component and what mechanisms may have produced it. 

Throughout this work we will refer to the galaxy's \lapex central\rapex stellar component as putative bulge (hereafter PB), an unbiased title that reflects our ignorance about its nature: it may be an off-set pseudo-bulge; it may have been formed in-situ, in an asymmetric burst of star formation; it may be the remnant of an external accreted dwarf galaxy; it may be an external galaxy crossing the line of sight (\citetalias{Rownd1994}; \citealt{Mihos2013}), bound or unbound to NGC 5474. Any of these explanations would require to call the PB with a different name. According to \cite{Fisher2010} this component has properties more similar to a pseudo-bulge than to a classical bulge (see \citealt{Kormendy2004,Fisher2008}), while \citet[hereafter B20]{Bellazzini2020} showed that its structural parameters are also consistent with the scaling relations of dwarf galaxies. For instance, the PB is similar to the dwarf elliptical galaxy (dE) NGC 205 in terms of stellar mass, V-band absolute magnitude and projected half-mass radius (\citetalias{Rownd1994}; \citealt{McConnachie2012}; \citetalias{Bellazzini2020}). \citetalias{Bellazzini2020} showed that the stellar populations of the PB are very similar to those dominating the stellar mass budget in the disc of NGC 5474 and the color-magnitude diagrams are fully compatible with systems lying at similar distance from us. Moreover, they constrained the maximum difference in radial velocity between the disc of NGC 5474 and the PB to $\sim50\kms$. Hence, if the PB is not a substructure of NGC 5474 it should be a satellite of it or, at least, another member of the M 101 group.

The presence of a stellar over-density to the South-West of the PB, a structure that extends for almost $1\amin^2$, adds up to NGC 5474 oddities \citepalias[see also Fig.~\ref{fig:ngc5474}]{Bellazzini2020}. The main population dominating the over-density is older than $2\Gyr$, very similar to the stellar population dominating the PB. Young stars in NGC 5474 appear not to be correlated to the over-density, and dominate instead the spiral arms, which extend $8\kpc$ out from the centre. The spiral pattern seen in optical is also marked by the $\HI$ distribution \citepalias{Rownd1994}. Among the possible explanations, it does seem plausible that such over-density may have been caused by a recent or on-going interaction between NGC 5474 and the off-centre PB, which may be also the cause of the galaxy's large scale, asymmetric recent star formation \citepalias{Bellazzini2020}.
 
In this work, we study the dynamical properties of NGC 5474 and investigate what the true nature of the PB may be by making use of realistic $N$-body hydrodynamical simulations, supported by analytic models. In Section~\ref{sec:setup} we describe the dynamical model built to match NGC 5474 and the PB: its properties, the observational data; the approach used to constrain the model. In Section~\ref{sec:model} we estimate the gravitational effects felt by NGC 5474 due to the PB to put limits on the large parameter space to be investigated with numerical simulations. In Sections~\ref{sec:equ} and \ref{sec:sims} we focus on simulations. After describing the method used to sample the initial conditions, we explore two different scenarios: i) a purely stellar system (without dark matter) orbiting within the plane of the disc of  NGC 5474, and ii) a compact early type dwarf galaxy (with its own dark-matter halo) moving on a polar orbit around NGC 5474. The latter case is obviously intended to explore the possibility that PB is a satellite of NGC 5474 that is seen near the centre of its disc only in projection. On the other hand, the former case is the mean by which we explore the effects of an off-centred bulge on the underlying gaseous and stellar disc. In particular we are interested to answer questions like: is the off-set position of the PB compatible with some kind of long-standing quasi-equilibrium configuration and/or a relatively regular rotation curve? Can a stellar PB resist the drag by dynamical friction and/or the tidal strain from the dark-matter halo of its parent galaxy? Is our understanding of the galaxy driven by projection effects? Could a possible interaction between NGC 5474 and the PB explain some of the other peculiarities of NGC 5474? Section~\ref{sec:concl} concludes.

\section{Setting the model}
\label{sec:setup}

Throughout this work we assume that the main structures that comprise the target are a dominant disk galaxy similar to NGC 5474 and the PB, a compact stellar component that can be either embedded in a dark-matter halo or not.

\subsection{NGC 5474}
\label{sec:modelscomp}

We describe NGC 5474 as a multi-component galaxy comprising a dark-matter halo, a gaseous disc and a stellar disc. We assume that the dark halo is spherical with \citet*[hereafter NFW]{NavarroFrenkWhite1996} density distribution
\begin{equation}\label{for:dm}
 \rhodm^{\rm NFW}(r) =\displaystyle\frac{4\rhos}{\displaystyle\left(\frac{r}{\rs}\right) \left(1+\frac{r}{\rs}\right)^2},
\end{equation}
where $\rs$ is the halo scale radius and $\rhos\equiv\rhodm^{\rm NFW}(\rs)$.

We assume that both the gaseous and the stellar discs are razor-thin exponential discs. The $\HI$ surface number density is given by
\begin{equation}\label{for:expgas}
 n(R) = \frac{\Mgas}{2\pi \hgas^2\mprot}\exp\biggl(-\frac{R}{\hgas}\biggr),
\end{equation}
where $\Mgas$ is the gas total mass, $\mprot$ is the proton mass (assuming, for simplicity, that the disc is fully composed by hydrogen atoms), and $\hgas$ is the $\HI$ disc scale length.

The stellar surface density is given by
\begin{equation}\label{for:expstar}
 \Sigmastar(R) = \frac{\Mstar}{2\pi \hstar^2}\exp\biggl(-\frac{R}{\hstar}\biggr),
\end{equation}
where $\Mstar$ is the disc total stellar mass and $\hstar$ is the stellar disc scale length. In our analysis, a model of NGC 5474 is fully determined by the free parameter vector $\bbxi \equiv \{\rhos,\rs,\hgas,\Mgas,\hstar,\Mstar\}$. To determine $\bbxi$, we fit a dataset of observations of NGC 5474 with our galaxy model. We anticipate that, in the subsequent analysis of NGC 5474 through hydrodynamical $N$-body simulations, we will drop the approximation of razor-thin discs for gas and stars in favour of realistic discs with non-negligible thickness.
 
\subsubsection{The dataset}
\label{sec:hd}

The first $\HI$ observations from \cite{vanderHulst1979} have too low velocity resolution ($\sim27\kms$) to allow for any detailed kinematic study. The $\HI$ data collected through VLA observations by \citetalias{Rownd1994} provide an $\HI$ velocity field relatively smooth and symmetric in the galaxy's central, unwarped region, with a rotation curve peaking at $\sim14\kms$. This is inconsistent with \citetalias{Kornreich2000} whose rotation curve is $7\kms$ systematically lower, even though the authors use the very same $\HI$ observations. Also, both $\HI$ rotation curves hardly agree with the $\Halpha$ emission, tracing the gas kinematics of the innermost galaxy's $3\kpc$ region. According to \citetalias{Epinat2008}, the $\Halpha$ rotation curve sharply rises up to $22\kms$ at $R\simeq2\kpc$ and it is barely consistent with \citetalias{Rownd1994} further out. Nonetheless, all the aforementioned studies agree and report the same off-set between the PB and the $\HI$/$\Halpha$ kinematic centres. For our study we rely on the rotation curve of \citetalias{Rownd1994} which provides a detailed description of the data reduction and has a large radial coverage. 


The dataset used to constrain the NGC 5474 galaxy model then consists of: i) the $\HI$ rotation curves derived starting from the analysis of \citetalias{Rownd1994} for the approaching and receding arms using a tilted ring model; ii) the observed $\HI$ column density profile of \citetalias{Rownd1994}; iii) the stellar disc parameters resulting from the stellar disc/bulge decomposition of \cite{Fisher2010}. 

\citetalias{Rownd1994} provides a collection of $\Nv=18$ points $\{\Rk,\vk^{\rm a}, \vk^{\rm r},\vk^{\rm t}\}$, with $k=0,...,\Nv$, where $\Rk$ is the distance of the observed point from the $\HI$ disc's kinematic centre, $\vk^{\rm a}$ ($\vk^{\rm r}$) is the corresponding velocity of the approaching (receding) arm as a result of the fit with the tilted ring model using half ring, while $\vk^{\rm t}$ is obtained using a complete ring. To rederive the $\HI$ rotation curve and determine a reliable error $\delta\vk$, accounting for the asymmetries of the two arms and the uncertain and low inclination we proceed as follows. For each radial bin $k$
\begin{itemize}
 \item[i)] we compute $\sigmak\equiv|\vk^{\rm a}-\vk^{\rm r}|/2$, as a measure of the velocity asymmetry between the two arms.
 \item[ii)] With a Monte Carlo approach, we sample $M=20000$ new velocities $\vk^j$, with $j=1,...,M$, from a Gaussian distribution with mean and standard deviation equal to $\vk^{\rm t}$ and $\sigmak$, respectively.
 \item[iii)] Each velocity $\vk^j$ is deprojected assuming a different inclination $i$, drawn from a uniform distribution over the range of inclinations  $[17^{\circ},25^{\circ}]$, consistent with the estimate of \citetalias{Rownd1994}.
 
 
 \item[iv)] We now have $M$ realizations of the intrinsic rotation velocity which we use to build the probability distribution of the rotation velocity of the considered bin. For each bin, we adopt the $50^\tth$ percentile of the distribution as rotation velocity and the minimum distance between the $84^{\tth}$ and $50^{\tth}$, and the $50^{\tth}$ and $16^{\tth}$ percentiles as error.
\end{itemize}

\begin{figure}
 \centering
 \includegraphics[width=.9\hsize]{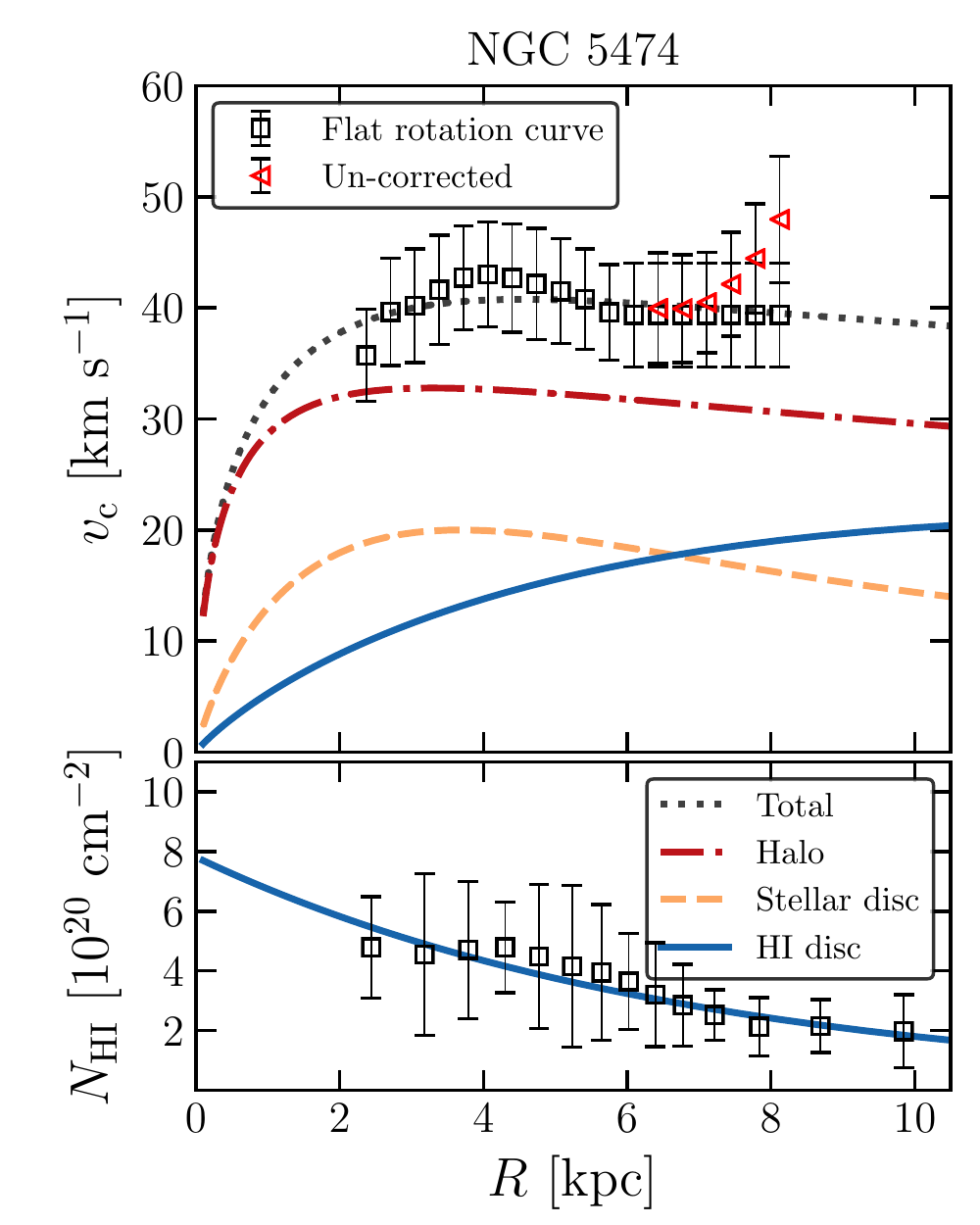}
 \caption{Top panel: $\HI$ rotation curve derived in Section~\ref{sec:hd} (squares with error bars), superimposed to the model rotation curve. We mark with different colors and line types the contributions to the total circular speed (black dotted curve) of the halo (red dash-dotted curve), the stellar disc (orange dashed curve) and the gaseous disc (blue solid curve). For comparison, we show also the uncorrected rotation curve (triangles with error bars). Bottom panel: $\HI$ observed column density profile as derived in Section \ref{sec:hd} (squares with error bars) superimposed to the model profile (blue solid curve, equation \ref{for:expgas}).}\label{fig:modelvsdata}
\end{figure}

The upper panel of Fig.~\ref{fig:modelvsdata} shows the $\HI$ rotation curve that we obtained. We mark the separate contributions of our reference NGC 5474 galaxy model with different colors, as we shall discuss in details in the following sections. At least for $R<8.5\kpc$, the rotation curves of the two arms from \citetalias{Rownd1994} are quite similar, with $\sigma\simeq2\kms$, so the velocity distribution is relatively smooth and symmetric. The deprojected rotation curve peaks at $R\sim3-4\kpc$, where $v\simeq42\kms$, and it slowly decreases out to $6\kpc$. For distances larger than $6\kpc$, the rotation curve rises up to $v\simeq50\kms$. As pointed out by \citetalias{Rownd1994}, the rise in the outermost regions is likely due to the $\HI$ disc's warp. We have no reason to believe that it would rise again in the outer parts, especially if this is a warped region. Since we are interested in getting a well motivated dynamical mass for NGC 5474, which can be used as a starting point for our simulations, we impose to our derived rotation curve to remain flat beyond $6\kpc$.

The $\HI$ column density profile has been derived from the $\HI$ column density map of \citetalias{Rownd1994}. We first plot the column density of each pixel as a function of the distance from the disc's kinematic centre. We build 15 radial bins, each containing the same number of pixels (approximately 100). For each bin, we compute the distribution of the $\HI$ surface density. We take the $50^{\tth}$ percentile of the distribution as a measure of the bin's $\HI$ column density, and the maximum interval between the $95^{\tth}$ and $50^{\tth}$, and $50^{\tth}$ and $5^{\tth}$ percentiles as error. We assume that the disc is razor-thin, and correct for the inclination according to the relation $\nint = \nobs \cos i$, where $\nint$ and $\nobs$ are, respectively, the intrinsic and observed surface density, $i=21^{\circ}$ (the mean inclination of the range used to estimate the rotation curve). We exclude the innermost bin, corresponding to a galactocentric distance of $R=1\kpc$, for which we do not have any kinematic information. At the end of the procedure, the $\HI$ column density profile consists of $\Ns=14$ triplets $\{\Ri,\nnk,\delta\nnk\}$, with $k=0,...,\Ns$ , which denote, respectively, the galactocentric distance ($\Rk$), the observed profile ($\nnk$) and the associated error ($\delta\nnk$). 

To derive a motivated range of stellar scale lengths and masses, we start from the stellar disc/bulge decomposition of \cite{Fisher2010}, who find that the best fit stellar disc model has $\logten\hstar/\kpc=3.09\pm0.06$, assuming a distance $d=5.03\Mpc$\footnote{\cite{Fisher2010} model the stellar disc of NGC 5474 with a razor-thin disc model, as in equation (\ref{for:expstar}).}. This value, converted to our distance $d=6.98\Mpc$ \citep{Tully2013}, gives $\hstar=1.71\pm0.23\kpc$. The estimate of \cite{Fisher2010} is based on data in the 3.6 $\mu$m band, a good tracer of stellar mass, with a weak dependence on age and metallicity. \citetalias{Bellazzini2020} found that the bulk of the stellar mass in the disc should be provided by intermediate to old age populations. According to the theoretical models by \cite{Rock2015}, the 3.6 $\mu$m mass-to-light ratio $\Upsilon$ for a Kroupa IMF and in the metallicity range relevant for NGC 5474 (Z$\in[0.0006,0.006]$; \citetalias{Bellazzini2020}) is $\Upsilon\sim0.5$ for a $5\Gyr$ old population and $\Upsilon\sim0.8$ for a $10\Gyr$ old population. In the following we take these two values as our reference.


Starting from the absolute magnitude in the 3.6 $\mu$m band $\Magn^{\disc}_{3.6}=-9.18\pm0.24$ as in \cite{Fisher2010}, for the adopted distance $d=6.98\Mpc$ and assuming $\Magn_{3.6,\odot}=2.24$ from \cite{Oh2008}, we convert the disc stellar luminosity into mass obtaining
\begin{equation}\begin{split}
    & \Mstar = 2.96\pm0.87\times10^8\Msun\quad \text{and} \\
    & \Mstar = 4.73\pm1.39\times10^8\Msun,
\end{split} \end{equation}
for $\Upsilon\sim0.5$ and $\Upsilon\sim0.8$, respectively. We conclude that, for the stellar disc, a plausible range of total stellar mass, accounting for all the uncertainties in the estimates provided above, is $\Mstar\in[2.1,6.1]\times10^8\Msun$, given a stellar disc scale length of $\hstar=1.71\pm0.23\kpc$. We note that our estimate of $\Mstar$ is consistent with \cite{Skibba2011}, who measure a total stellar mass $\Mstar=5\times10^8\Msun$ using far-infrared imaging from the Herschel Space Observatory for galaxies in the KINGFISH project.

\subsubsection{Fitting procedure}
\label{sec:method}

The log-likelihood of a galaxy model $\ln\LL(\bbxi|\data)$, defined by the parameter vector $\bbxi$, given the data $\data$ is

\begin{equation}\label{for:chi2}
 \ln\LL = \ln\LL_v + \ln\LL_n .
\end{equation}
The first term in the r.h.s. is
\begin{equation}\label{for:chiv}
 \ln\LL_v = -\frac{1}{2}\sum_{k=0}^{\Nv}\biggl(\frac{\vc(\Rk;\bbxi) - \vk}
 {\delta \vk}\biggr)^2,
\end{equation}
where $\vc$ is the model circular speed given by
\begin{equation}\label{for:vv}
 \vc^2 = \vh^2 + \vgas^2 + \vstar^2,
\end{equation}
with 
\begin{equation}\label{for:vh}
 \vh^2\equiv 8\pi\rs\rhos G\biggl[\frac{\ln(1+r/\rs)}{r/\rs} - \frac{1}{1+r/\rs}\biggr]
\end{equation}
the contribution to the circular speed due to the halo, and 
\begin{equation}\begin{split}\label{for:vd}
 & \vany^2 =  \frac{2G\Many}{\hany}\yany^2\times \\ &  [I_0(\yany)K_0(\yany) - I_1(\yany)K_1(\yany)]
\end{split}\end{equation}
the contribution to the circular speed of any of the discs, measured in the discs plane, with $\comp=\star$ for the stellar disc and $\comp=\gas$ for the gaseous disc. In equations (\ref{for:vh}) and (\ref{for:vd}), $G$ is the gravitational constant and $I_n$ and $K_n$ are Bessel's functions of the $n$-th order. In equation (\ref{for:chi2}), the latter term is
\begin{equation}\label{for:chiS}
 \ln\LL_n = -  \frac{1}{2}\sum_{k=0}^{\Ns}\biggl(\frac{n(\Rk;\bbxi) - \nnk}
 {\delta \nnk}\biggr)^2,
\end{equation}
where $n$ and $\nnk$ are the model (equation \ref{for:expgas}) and observed $\HI$ surface density profiles. Since the discs' and halo contributions may be highly degenerate, we fix the stellar disc parameters to observationally motivated values, consistent with the aforementioned estimates. We adopt  $\hstar=1.71\kpc$ as stellar disc scale length and the average value $\Mstar=4.1\times10^8\Msun$ as stellar mass.

\begin{figure*}
 \centering
 \includegraphics[width=1\hsize]{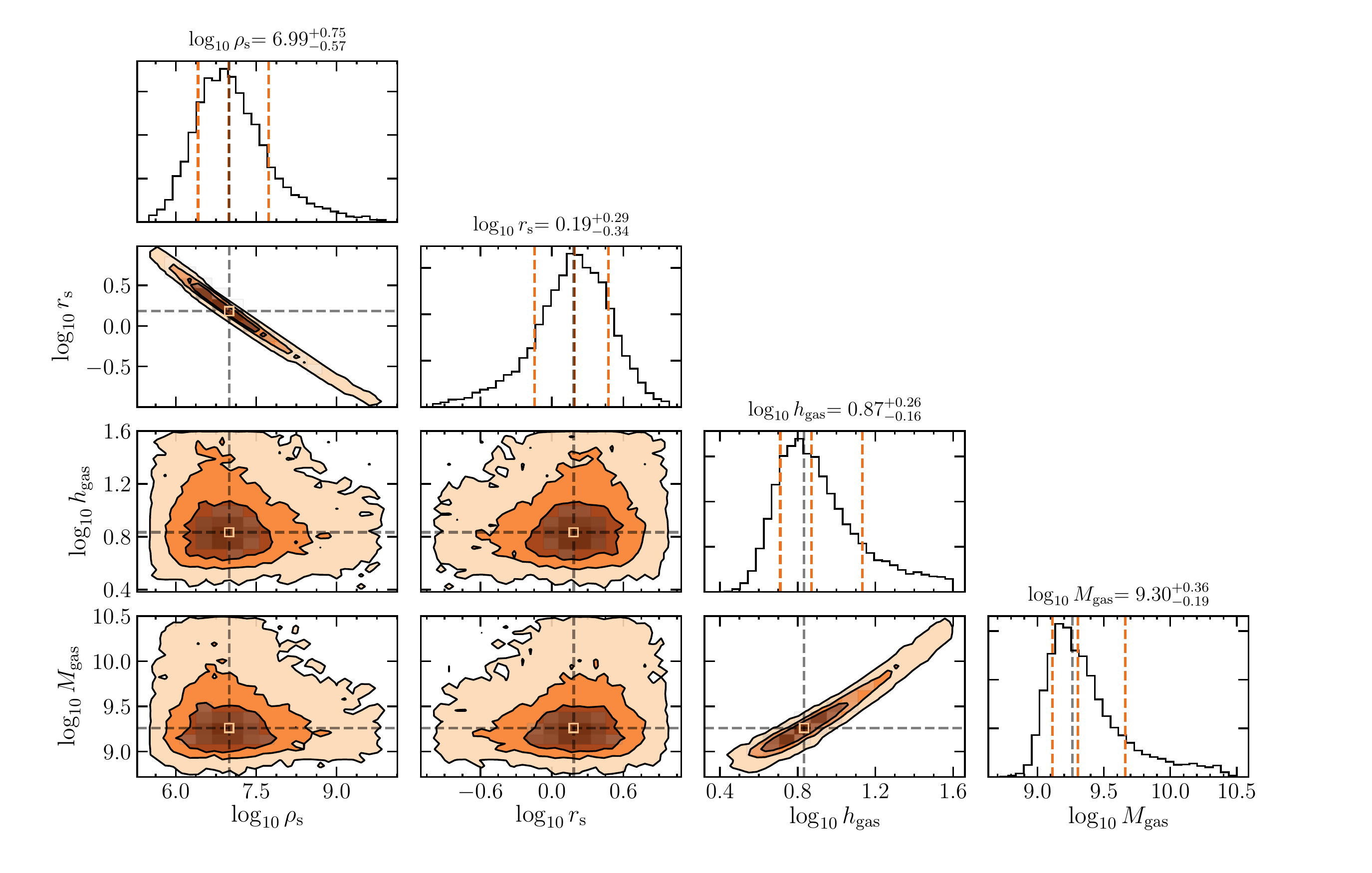}
 \caption{One- and two-dimensional marginalized posterior distributions over the model free parameters $(\rhos,\rs,\hgas,\Mgas)$. The black curves in the two-dimensional marginalized distributions correspond to regions enclosing, respectively, 68\%, 95\% and 99\% of the total probability. The orange vertical lines in the one-dimensional marginalized distributions correspond to the 16$^{\tth}$, 50$^{\tth}$ and 84$^{\tth}$ percentiles, used to estimate the uncertainties over the models free parameters. The vertical grey lines in the marginalized one-dimensional distributions, and the  squares in the marginalized two-dimensional distributions mark the position of the reference model (see also Table~\ref{tab:param}).}\label{fig:posterior}
\end{figure*}

\begin{table*}
 \begin{center}
  \caption{Main parameters of the analytic NGC 5474 models of halo, $\HI$ and stellar discs. $\rhos$ and $\rs$: reference density and scale radius of the NFW dark-matter density profile (equation~\ref{for:dm}); $\Mgas$: total mass of the $\HI$ disc; $\hgas$: $\HI$ disc's scale length (equation~\ref{for:expgas}); $\Mstar$: total mass of the stellar disc; $\hstar$: stellar disc scale length (equation~\ref{for:expstar}). The parameters ($\rhos,\rs,\hgas,\Mgas$) are determined as described in Section~\ref{sec:method}, while we fixed the parameters of the stellar disc to the ones derived in Section~\ref{sec:hd}. The middle row lists the $1\sigma$ uncertenties over the model free parameters, while the bottom row lists the parameters of the reference model adopted throughout this work.}\label{tab:param}
  \begin{tabular}{lcccccc}
   \hline\hline
      NGC 5474 & \multicolumn{2}{c}{DM halo} & \multicolumn{2}{c}{HI disc} & \multicolumn{2}{c}{Stellar disc} \\
   \hline\hline
    parameter & $\rhos$ [$10^6\Msun\kpc^{-3}$]	&	$\rs$ [$\kpc$] 	&    $\Mgas$ [$10^9\Msun$] & $\hgas$ [$\kpc$]	& $\Mstar$ [$10^8\Msun$] 	&	$\hstar$ [$\kpc$] \\
   value & $9.77^{+45.18}_{-7.14}$		&	$1.55^{+1.47}_{-0.84}$	&    $2.00^{+2.58}_{-0.71}$ 	& 	$7.41^{+6.08}_{-2.28}$		 &     $[2.1,6.1]$ & $1.71^{+0.23}_{-0.23}$ \\
   reference model & $9.95$				&	$1.51$			&    $1.82$			& 	$6.76$	& $4.1$ & $1.71$			\\
   \hline\hline
  \end{tabular}
 \end{center}
\end{table*}

\begin{table*}
 \begin{center}
  \caption{Main parameters of the PB relevant for this work. $\Reff$: PB effective radius from \citetalias{Bellazzini2020}; $\Mb$: PB total stellar mass as  derived in Section~\ref{sec:PB}; $m$: PB S\'{e}rsic index (equation \ref{for:ser1}) from \citetalias{Bellazzini2020};  $\rhosPB/\Mb$: PB dark-matter halo scale density-to-stellar mass (equation \ref{for:PBhalo}); $\rsPB$ and $\rtPB$: PB dark-matter halo scale and truncation radii, respectively (see equation \ref{for:PBhalo}).}\label{tab:paramPB}
  \begin{tabular}{lcccccc}
   \hline\hline
    Putative bulge & \multicolumn{3}{c}{Stars}         & \multicolumn{3}{c}{Dark-matter }          \\
   \hline\hline
    parameter   &   $\Reff$ [$\kpc$]	& $\Mb$ [$10^8\Msun$]	& 	$m$ 		 & $\rhosPB/\Mb$ [$\kpc^{-3}]$  & $\rsPB$ [$\kpc$] &  $\rtPB$ $[\kpc$]    \\
    value       &   $0.484$     	    & $[0.5,2]$		        &	$0.79$ & $5.30\times10^6$        &  2.5	 & 15\\
   \hline\hline
  \end{tabular}
 \end{center}
\end{table*}

We perform a parameter space search using a Markov Chain Monte Carlo (MCMC) method. We run 16 chains, each evolved for 7000 steps, using a classical Metropolis-Hastings sampler \citep{Metropolis1953,Hastings1970} to sample from the posterior. We adopt flat priors on the free parameters. After a burn-in of 3000 steps (which we eliminate as a conservative choice), we use the remaining steps to build the posterior distributions over $\bbxi$. Fig.~\ref{fig:posterior} shows the marginalized one- and two-dimensional posterior distributions over the models' parameters. We estimate the uncertainties on the models' free parameters using the 16$^{\tth}$, 50$^{\tth}$ and 84$^{\tth}$ percentiles of the corresponding marginalized one-dimensional distributions.

We note that since the PB is off-set with respect to the disc kinematic centre we cannot include it in our axisymmetric model of the rotation curve. However, after computing the PB mass from its stellar population, in Section~\ref{sec:model} we try to estimate the possible effects of the PB on the gas kinematics.

Figure~\ref{fig:modelvsdata} shows the newly derived $\HI$ rotation curve superimposed to that of the model. We highlight with different colors the contributions of the different components. The dark-matter halo dominates over the stellar and gaseous components at all the radii covered by the kinematic data. Even if the total stellar mass is lower than the total $\HI$ mass, in the central regions the stars provide a significant contribution, dominant over the $\HI$ disc for $R\le6\kpc$, due to the very different gas and stellar scale lengths. As a reference, we estimate $\Mstar/\Mgas|_{2\kpc}\simeq 2$  within $2\kpc$, and $\Mstar/\Mgas|_{9\kpc}\simeq0.55$ at the larger distance $9\kpc$. Also, we measure an $\HI$ mass $\Mgas|_{9\kpc}\simeq7\times10^7\Msun$ within $R=9\kpc$, consistent with the estimate of \citetalias{Rownd1994}. The bottom panel of Fig.~\ref{fig:modelvsdata} shows the $\HI$ disc column density as a function of the galactocentric distance, superimposed to the model. \citetalias{Rownd1994} reports that the disc surface density flattens in the central parts, which is consistent with an exponential disc model with a large scale length (equation~\ref{for:expgas}, Fig.~\ref{fig:posterior}). If the inner $\HI$ surface density were constant and equal to the innermost point of the observed profile, our exponential model would overestimate the $\HI$ mass within $2\kpc$ by only 10\%, which, given the uncertainties on the observed profile and the model assumptions, we consider of negligible impact.

We define our reference model as the model (i.e. a set of $\bxi$) with the maximum likelihood (equation \ref{for:chi2}) a posteriori. Table~\ref{tab:param} lists the main parameters of disc and halo here derived, and the parameters of the reference NGC 5474 model.

\subsection{Putative bulge}
\label{sec:PB}

In our analysis the PB can be either embedded in a dark-matter halo or not. We represent its stellar component with a spherical \cite{Sersic1968} model, whose surface brightness profile is
\begin{equation}\label{for:ser1}
 I(R)=\Ie\exp\biggl[-\bm\biggl(\frac{R}{\Reff}\biggr)^{\frac{1}{m}}\biggr],
\end{equation}
where
\begin{equation}\label{for:ser2}
 \Ie = \frac{b^{2m}}{2\pi m\Gamma(2m)}\frac{\Ltot}{\Reff^2}.
\end{equation}
In equations (\ref{for:ser1}) and (\ref{for:ser2}), $\Gamma$ is the Gamma function, $\Ltot$ is the total PB luminosity, $m$ is the S\'{e}rsic index, related to $\bm$ as in equation 18 of \cite{Ciotti1999}, and $\Reff$ the effective radius (i.e. the distance on the plane of the sky from the PB's centre that contains half of the total PB's luminosity $\Ltot$).

We adopt $m=0.79$ and $\Reff=0.484\kpc$ from \citetalias{Bellazzini2020} (see also Table~\ref{tab:paramPB}) and we infer the stellar mass from the total 3.6$\mu$m luminosity by \cite{Fisher2010} using the same mass-to-light ratios we adopted for the disc. The absolute magnitude in the 3.6$\mu$m band estimated by \cite{Fisher2010} is $\Magn_{3.6}=-16.44\pm0.22$, which, converted assuming a distance $d=6.98\Mpc$, gives $\Magn_{3.6}=-17.15\pm0.22$. We follow \cite{Forbes2017} and, assuming the $1\sigma$ limits in the luminosity and adopting $\Magn_{3.6,\odot}=2.24$ from \cite{Oh2008}, we get 
\begin{equation}\begin{split}
 & \Mb=0.70^{+0.18}_{-0.12}\times10^8\Msun \quad\text{and} \\
 & \Mb=1.14^{+0.26}_{-0.21}\times10^8\Msun,
\end{split}\end{equation}
for $\Upsilon=0.5$ and $\Upsilon=0.8$, respectively\footnote{As for the stellar disc, the mass-to-light ratios $\Upsilon$ are in the 3.6 $\mu$m band.}. Based on the aforementioned estimates, we take the slightly wider $\Mb\in[0.5,2]\times10^8\Msun$ as a reasonable range of stellar masses to explore for the PB. The main parameters $(m,\Reff,\Mb)$ relevant to this work are listed in Table~\ref{tab:paramPB}. At any rate, we should recall that any derivation of the PB and disc masses from $\Upsilon$ depends on the adopted IMF, and that switching, e.g., from a Kroupa to a Salpeter's IMF can change the resulting masses by a factor of 3 (see Fig.s 13 and 14 of \citealt{Rock2015}).

When present, the PB dark-matter halo has density distribution
\begin{equation}\label{for:PBhalo}
 \rhodmPB(r) = \displaystyle\frac{\rhosPB}{\displaystyle\frac{r}{\rsPB}\left(1+\displaystyle\frac{r}{\rsPB}\right)^2}e^{-\left(\displaystyle\frac{r}{\rtPB}\right)^2}
\end{equation}
i.e., a truncated NFW model, where $\rhosPB$ and $\rsPB$ are, respectively, the halo scale density and the characteristic radius, while $\rtPB$ is the halo truncation radius. 

According to estimates of the stellar-to-halo mass relation \citep{Read2017}, for a galaxy with stellar mass $\Mb\in[0.5,2]\times10^8\Msun$, one would expect a virial-to-stellar-mass ratio $\MvirPB/\Mb\simeq100$. However, since we consider the PB as an external galaxy orbiting around NGC 5474, we expect the PB halo to be less massive than estimated and to be truncated well before its nominal virial radius because of tidal interactions. Also, the structural properties of its stellar component resemble the ones of a typical dE galaxy (see also \citetalias{Bellazzini2020}), which we do not expect to be significantly dominated by dark matter \citep{McConnachie2012}. As such, we use the above $\MvirPB/\Mb$ only as a reference value to estimate $\rsPB$. Adopting the halo mass-concentration relation from \cite{MunozCuartas2011}, from which we estimate a concentration $\logten c=1.19-1.22$, we get a halo scale radius $\rsPB=2.17-2.91\kpc$. We set $\rsPB=2.5\kpc$, but we do not expect our results to depend substantially on $\rsPB$. To derive a new mass scale we impose that the dynamical-to-stellar-mass ratio $\MdynPB/\Mb$, evaluated at the stellar half-mass radius $\rh$, is
\begin{equation}\label{for:mdynmb}
\frac{\MdynPB}{\Mb}\biggr|_{\rh} \sim 2,
\end{equation}
which is approximately the ratio expected for a dE of sizes and structure similar to the PB \citep{McConnachie2012}. We truncate the PB halo at $\rtPB=15\kpc$. As we will later discuss, such a value is slightly less than the initial distance we set between the PB and the NGC 5474 centres in the simulations of Section~\ref{sec:resPBDM} and it avoids the halo PB to be unreasonably massive. With these choices, the PB has a dark-matter halo 20 times as massive as its stellar component. As a reference, the most massive PB halo, corresponding to a stellar mass $\Mb=2\times10^8\Msun$, has a total dark-matter mass $\simeq4\times10^9\Msun$, approximately the same as the halo virial mass of NGC 5474. Table \ref{tab:paramPB} lists the relevant parameters of the PB used throughout this work.

\section{Constraints from observations}
\label{sec:model}

\begin{figure*}
 \centering
 \includegraphics[width=.85\hsize]{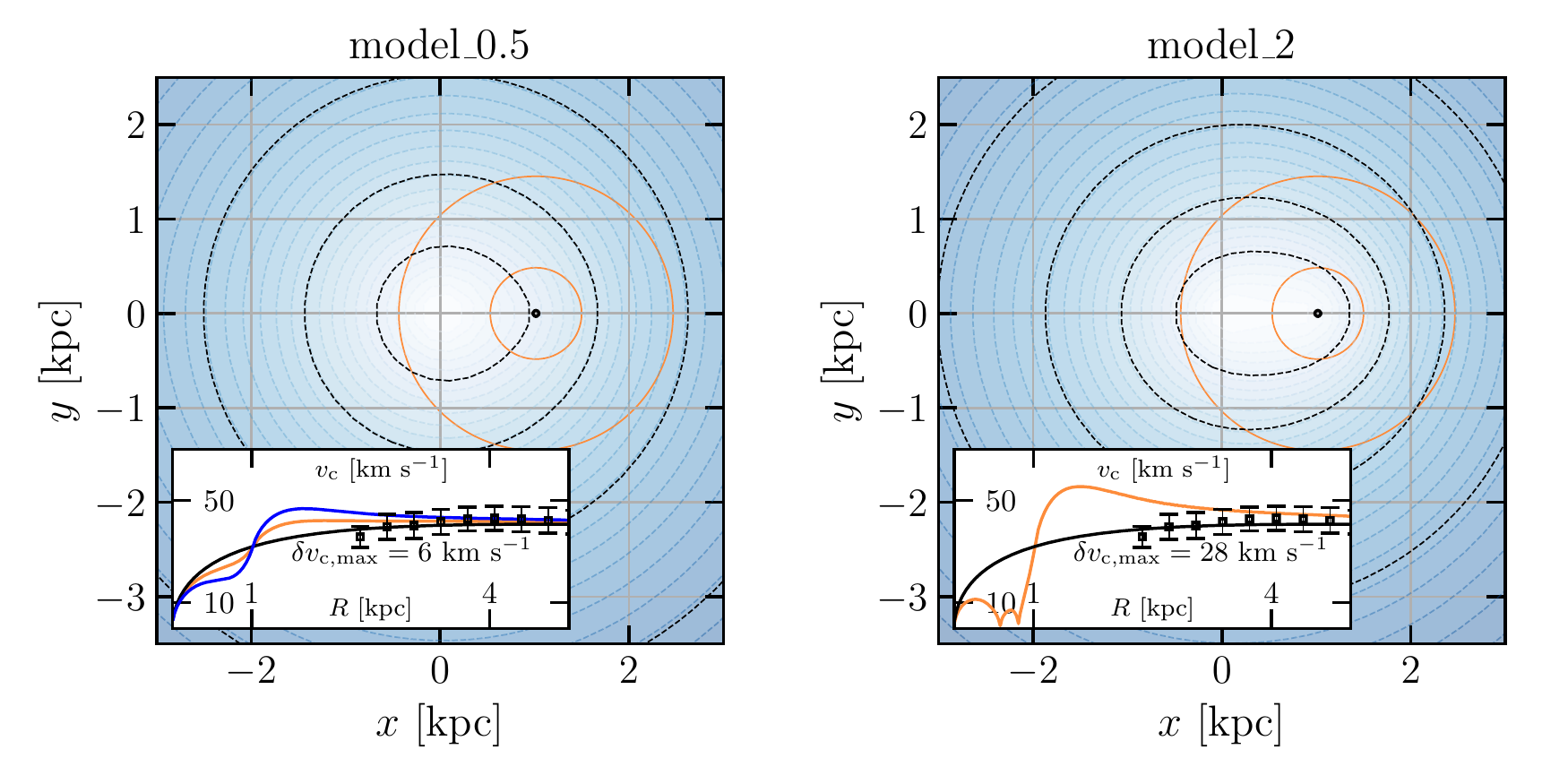}
 \caption{Left-hand panel: total (halo, stellar disc, $\HI$ disc and PB) gravitational potential map in the discs' equatorial plane when $\Mb=0.5 \times10^8\Msun$ (model\_0.5).  The shades of blue, from light to dark, mark regions of increasing potential. The PB is located at $(x,y)=(1\kpc,0)$ (black point). The orange circles show distances corresponding to $\Reff$ (inner) and $3\Reff$ (outer), with $\Reff$ the PB's effective radius (Table~\ref{tab:paramPB}). Small inset at the bottom of the left panel: \lapex circular speed\rapex as a function of the distance from the $\HI$ disc's kinematic centre of NGC 5474 without (black curve) and with (orange curve) the PB. In the latter case the circular speed is computed following equation (\ref{for:vc}) along the principal axis $y=0$. The black points with errorbars show the rotation curve derived in Section~\ref{sec:hd}. Right-hand panels: same as the left-hand panels, but for $\Mb=2\times10^8\Msun$ (model\_2). As a further comparison, in the inset of the left-hand panel we also show the circular speed obtained in the intermediate case of a PB with $\Mb=10^8\Msun$ placed $1\kpc$ away from the galaxy kinematic center (blue curve).}\label{fig:potmap}
\end{figure*}

By means of the analytic model derived in Sections~\ref{sec:modelscomp} and \ref{sec:PB}, we quantify the mutual effects that the galaxy and the PB may have on each other when the latter is placed within the discs' equatorial plane. This allows us to put further constraints on the large parameter space we will explore with hydrodynamic $N$-body simulations in the following sections.

\subsection{Effects of the presence of the PB}
We start analyzing the two scenarios of a PB with $\Mb=0.5\times10^8\Msun$ and $\Mb=2\times10^8\Msun$, respectively the lower and upper limits of the mass range derived in Section~\ref{sec:PB}. We assume no dark matter since we do not consider the PB as an external galaxy, and we first place it on the discs' plane, $1\kpc$ away from the kinematic centre, with its current size and \Sersic index ($\Reff=0.484\kpc$, $m=0.79$, see Table~\ref{tab:paramPB}). We refer to the two analytic models as, respectively, model\_0.5 and model\_2, where the number indicates the PB mass, in units of $10^8\Msun$. Using these models, we roughly estimate the minimum and maximum distortions we may expect to see in the $\HI$ disc's velocity field due to the presence of the PB on plane.

Figure~\ref{fig:potmap} shows the total gravitational potential map of model\_0.5 (left panel) and model\_2 (right panel) on a portion of the equatorial plane. The total gravitational potential has been computed summing the separate contributions of the discs, halo and PB. The PB of model\_2 contributes to the total gravitational potential so intensely that the potential well shifts towards the PB centre. In this circumstance, it is hard to imagine an equilibrium configuration in which the PB and the $\HI$ disc kinematic centre are off-set.


A different perspective is given by the small insets of Fig.~\ref{fig:potmap}, where we show the circular speed of the NGC 5474 model (discs and halo), superimposed to the circular speed computed from model\_0.5 (left panel) and model\_2 (right panel). Of course, these systems have lost their cylindrical symmetry, so the concept of circular speed makes no strict sense, but, at least in model\_0.5 where the PB contribution to the potential is sub-dominant, this exercise still helps to quantify the magnitude of the $\HI$ disc's velocity field perturbations. Calling the $(x,y)$-plane the equatorial plane, and $\Phitot$ the model's total gravitational potential, we define the \lapex circular speed\rapex
\begin{equation}\label{for:vc} 
 \vc \equiv \sqrt{x\biggl|\frac{\DD\Phitot}{\DD x}\biggr|_{y=0}}.
\end{equation}
The latter is computed for the models with and without the PB, along the line $y=0$, where the $y=0$ axis is aligned with the PB centre, when present. Then we measure $\delta\vcmax$, i.e. the maximum difference in circular speed between the models with and without the PB, in a region of approximately $3\Reff\simeq1.5\kpc$ around the PB's centre. 

In model\_2 (Fig.~\ref{fig:potmap}, right panel) the PB produces distortions as high as $\delta\vcmax\simeq28\kms$, which is not even consistent with the rotation curve derived in Section~\ref{sec:hd} from the $\HI$ kinematics. This is not surprising since the PB centre is very close to the minimum value of the gravitational potential. While it is highly unlikely that the PB as in model\_2 can be located onto the $\HI$ disc, given the large distortion it generates over the wide area covering approximately $3\Reff$, this is not excluded for model\_0.5, especially due to the lack of kinematic information within $R\sim2\kpc$ (Fig.s~\ref{fig:modelvsdata} and \ref{fig:potmap}). 
In the inset in the left panel of Fig.~\ref{fig:potmap} we also show the circular speed obtained when we place a PB with $\Mb=10^8\Msun$ $1\kpc$ away from the galaxy kinematic centre. The maximum distortions the PB generates in this case are as high as $\delta\vcmax\simeq11\kms$, but still the overall profile is marginally consistent with the observed one within the errorbars.

On this basis, we will restrict the range of possible PB stellar masses to $\Mb\in[0.5,1]\times10^8\Msun$ in any further analysis, since we expect more massive PB to have critical effects on the $\HI$ disc of NGC 5474.


\subsection{The tidal radius of the PB}


We now focus on cases in which the PB moves along orbits co-planar with the galaxy discs. When on plane, we expect the dynamical friction and the tidal force field of NGC 5474 to be the main drivers of the PB evolution. While the former makes the PB sink towards the galaxy centre on a relatively short timescale (for details, see Section~\ref{sec:sim1pb}), the latter is responsible for the PB mass loss and the development of possible non-equilibrium features.

We quantify the effects of the tidal force field of NGC 5474 on the PB by computing its tidal radius $\rt$. Often the tidal radius $\rt$ is estimated as (\citealt{BinneyTremaine2008}, equation 8.91)
\begin{equation}\label{for:rt}
    \rt = \RPB\biggl(\frac{\Mb}{3\Mtot(\RPB)}\biggr)^{1/3},
\end{equation}
assuming that the size of the PB is negligible with respect to its distance from NGC 5474. In the above equation $\Mtot(\RPB)$ is the total mass of NGC 5474 enclosed within $\RPB$, and $\RPB$ is the distance of the PB from the centre of NGC 5474. For a PB with mass $\Mb=0.5\times10^8\Msun$ and $\Reff=0.484\kpc$, the tidal radius is $\rt\simeq2\rh$ at $\RPB=7\kpc$. Considering that the PB mass enclosed within $2\rh$ is $\simeq92\%$ of its total mass, we expect the tidal field of NGC 5474 to be of small impact on the PB structural properties at any $\RPB\gtrsim7\kpc$.

In this on-plane scenario it sounds then legitimate to take $\RPB=7\kpc$ as PB initial position and $\vc(\RPB)\simeq42\kms$ in the azimuthal direction as initial velocity, i.e. a circular orbit. Even if we considered a larger $\RPB$, dynamical friction would anyway make the PB sink within the disc eventually reaching $7\kpc$ with negligible or minimal mass loss because of a larger $\rt$. Whether the PB is the galaxy's pseudo-bulge, or it is the remnant of an external galaxy, or it has formed in-situ in a burst of star formation happened at least $2\Gyr$ ago (compatibly with its dominant stellar population, \citetalias{Bellazzini2020}), the observed, present-day PB would be the end-state of the orbital decay of any of these configurations.

Given that the systems are extended, to follow the evolution of the PB tidal radius from $\RPB=7\kpc$ to the central regions we rely on a more realistic approach. In a reference frame where the PB and the galaxy kinematic centre are aligned along the $x$-axis, we estimate the PB's tidal radius $\rt$ by computing the position $\bbx=(\rt<\RPB,0,0)$ (i.e. on the equatorial plane and along the $x$-axis), where the effective potential
\begin{equation}\label{for:rt2}
    \Phieff(\bbx) = \Phitot(\bbx) + \PhiPB(\bbx-\bbRPB) - \frac{1}{2}(\Omega\rt)^2
\end{equation}
has a saddle point. Here, $\bbRPB\equiv(\RPB,0,0)$ and $\Omega\equiv\vc(\RPB)/\RPB$, i.e. the angular speed obtained from the NGC 5474 model circular speed (Fig.~\ref{fig:modelvsdata}) at a distance $\RPB$. 

\begin{figure}
    \centering
    \includegraphics[width=1\hsize]{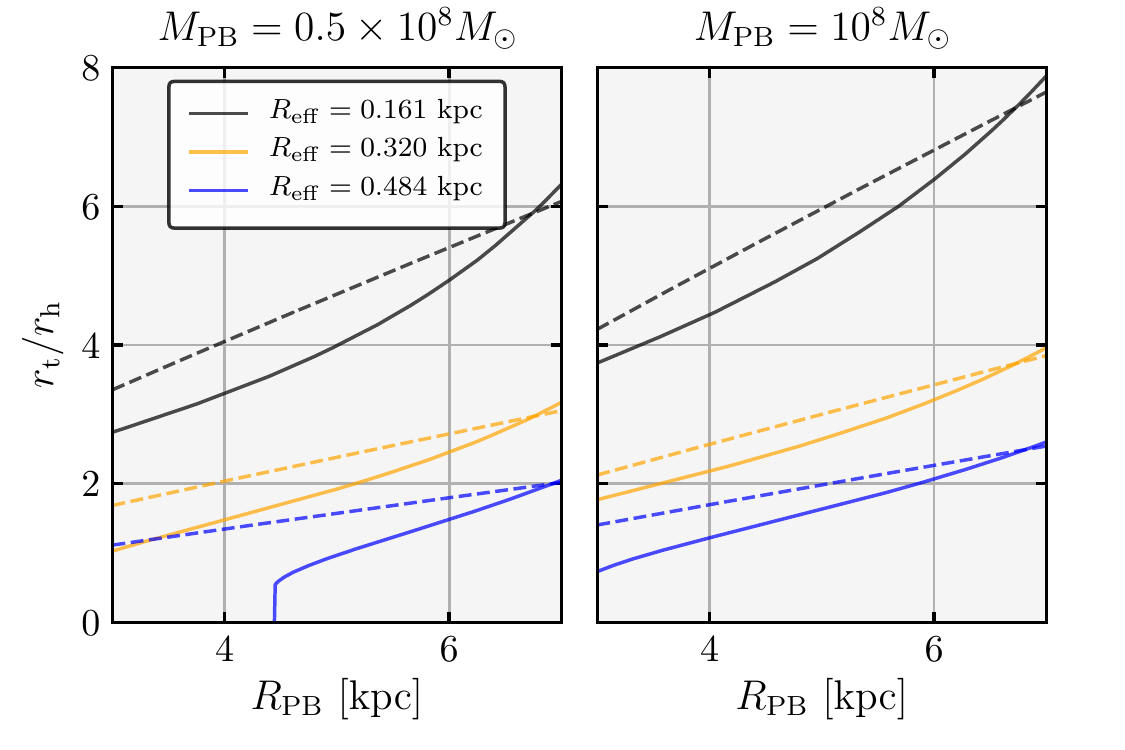}
    \caption{Left panel: ratio between tidal radius $\rt$ and the half-mass radius $\rh$ of a PB with total stellar mass $\Mb=0.5\times10^8\Msun$ (and no dark matter) as a function of the distance from the system's kinematic centre. The tidal radius is computed following equation \ref{for:rt} (dashed curves) and according to equation \ref{for:rt2} (solid curves). Each color refers to PBs with $\Reff=0.161\kpc$ (black curves), $\Reff=0.320\kpc$ (orange cuvers) $\Reff=0.484\kpc$ (purple curves). Right panel: same as the left panel, but for a PB with mass $\Mb=10^8\Msun$.}
    \label{fig:rt}
\end{figure}

Figure~\ref{fig:rt} shows the PB tidal radius as a function of the distance from the discs' centre. We compute $\rt$ according equation (\ref{for:rt2}) and, as a comparison, we also show $\rt$ computed according the classical (\ref{for:rt}). In addition to $\Mb=0.5\times10^8\Msun$, we examined the case in which $\Mb=10^8\Msun$. For each mass value, we considered three cases with $\Reff=0.484\kpc$, $\Reff=0.320\kpc$, and $\Reff=0.161\kpc$, because, as an effect of the tidal force field, the PB may become more extended (see, e.g. \citealt{Iorio2019}). We notice that equations~(\ref{for:rt}) and (\ref{for:rt2}) do not account for the PB mass loss, so even if $\rt$ decreases along the orbit, the PB total mass is the same as the initial one.

According to equation (\ref{for:rt2}), while the tidal radius of the less extended PBs is always larger than at least three half-mass radii, in the remaining cases the tidal radius shrinks fast to less than two PB half-mass radii at $3\kpc$. We note that the tidal radius computed as in its classical formulation (equation \ref{for:rt}) and as in equation (\ref{for:rt2}) gives approximately the same results when the PB is sufficiently far from the galaxy's centre, while equation (\ref{for:rt2}) provides an estimate of $\rt$ sensibly lower when the PB is close to the galaxy centre. For the less massive ($\Mb=0.5\times10^8\Msun$) and most extended PB ($\Reff=0.484\kpc$), the effective potential (\ref{for:rt2}) does not even have a saddle point, meaning that the truncation radius is formally zero. We do not show predictions for $\RPB<3\kpc$ since we expect the PB to have lost such a significant amount of mass to make ineffective also the use of equation (\ref{for:rt2}). At least for the most extended and least massive PB we may expect any effect due to the tidal force field of NGC 5474 to be extremely intense.

To conclude, on the grounds of these analytic models, it seems very unlikely that the PB would be as massive as $2\times10^8\Msun$, if placed on the discs' plane of NGC 5474. If that would be the case, we should be able to see strong distortions in the $\HI$ velocity field map that are, instead, unseen. Moreover, although we have considered the PB as only made of stars, we can interpret this result as an upper limit on the total PB dynamical mass since the effects on the $\HI$ disc on which we have focused are purely gravitational.

Even if we may expect the chance of survival of large off-centred PBs to be low, due to the galaxy's intense tidal field force, we cannot either exclude or prove that more favorable initial conditions, as a less extended initial PB, may produce the required off-set, a smooth $\HI$ velocity field and a regular PB spatial distribution. Starting from the predictions of the analytic models we just examined, we consider more quantitatively these scenarios through our hydrodynamic $N$-body simulations, which are described in the following sections.


\section{Set-up of the simulations}
\label{sec:equ}

\subsection{The \AREPO code}
\label{sec:arepo}

All our simulations are performed using the moving-mesh hydrodynamic code \AREPO \citep{Springel2010}, as implemented in its publicly-released version\footnote{\url{https://arepo-code.org/}.} \citep{Weinberger2020}. \AREPO combines the advantages of both Lagrangian smoothed particle hydrodynamics (SPH) and Eulerian hydrodynamics on an unstructured mesh with adaptive mesh refinement (AMR). The mesh is constructed from a Voronoi tessellation of a set of discrete points, used to solve the hyperbolic hydrodynamic equations with a finite-volume technique and it is free to move with the fluid flow. In our case, the mesh grid sizes are refined in such a way to ensure that each gas cell has approximately constant mass, allowing one to sample with a large number of cells high density regions. The gas moving mesh is coupled to a particle-mesh algorithm and oct-tree approach \citep{Barnes1986} to solve the Poisson equation and compute both collisional and collisionless particles accelerations.

\AREPO has been extensively employed to deal with a large number of astrophysical problems, such as AGN winds and feedback \citep{Costa2020}, stellar evolution and interstellar medium enrichment processes \citep{vandeVoort2020}, spiral arms formation mechanisms \citep{Smith2014}, and run state-of-the-art large volume cosmological simulations of galaxy formation such as the latest IllustrisTNG simulations \citep{Illustris2018a,Illustris2018b,Illustris2018c,Illustris2018d,Illustris2018e}, or zoom-in cosmological magneto-hydrodynamical simulations as the ones from the Auriga project \citep{Grand2017}. For a detailed review see \cite{Weinberger2020} and references therein.

\subsection{Realization of NGC 5474}
\label{sec:ic}

\subsubsection{Density distributions}
\label{sec:HDICs}

We sample the halo's and discs' initial conditions (hereafter ICs) following \cite{Springel2005}. The dark-matter halo density follows the \cite{Hernquist1990} model 
\begin{equation}\label{for:hern1}
\rhodm(r) = \frac{\Mtot}{2\pi}\frac{a}{r}\frac{1}{(r+a)^3},
\end{equation}
whose mass profile is given by
\begin{equation}\label{for:hern2}
\Mdm(r) = \Mtot\frac{r^2}{(a+r)^2},
\end{equation}
where $a$ is the Hernquist's scale radius and $\Mtot$ the halo total mass.

Dark-matter haloes are often represented with NFW models (equation \ref{for:dm}). However, the use of an Hernquist model (which has the same inner slope as the NFW model) is motivated by the fact that it has finite mass, and an analytic, ergodic distribution function (hereafter DF), which simplifies the sampling of the halo particle velocities. To link the Hernquist model to the NFW profile we require that the Hernquist total mass is equal to the NFW halo virial mass $\Mvir$ (i.e. the enclosed mass within the virial radius), and we impose that the two profiles share the same normalization in the central parts. As a consequence, provided, for instance, $\Mvir$ and $\rs$ for the NFW halo, the corresponding Hernquist halo is fixed with parameters
\begin{equation}\begin{split}\label{for:hern3}
 & \Mtot = \Mvir \\
 & a = \rs \sqrt{2[\ln(1+c)-c/(1+c)]}. 
\end{split}\end{equation}
(see Fig. 1 of \citealt{Springel2005})

The radial density profiles of the gaseous and stellar discs follow from equations (\ref{for:expgas}) and (\ref{for:expstar}), respectively, where for the gas we have switched from particle number density to mass density. We drop the razor-thin disc approximation and let the discs have a non-negligible thickness. For the stellar disc, the vertical profile stratifies with radially constant scale height $\zstar$, so that its full three dimensional density distribution is given by
\begin{equation}\label{for:fullstar}
    \rhostar(R,z) = \frac{\Mstar}{4\pi\hstar^2\zstar}e^{-\bigl(\frac{R}{\hstar}\bigr)}\sech^2\biggl({\frac{z}{\zstar}\biggr)}.
\end{equation}

The vertical profile of the $\HI$ disc is determined from the vertical hydrostatic equilibrium
\begin{equation}\label{for:idroeq}
 \frac{\DD\Phitot}{\DD z} = - \frac{1}{\rhogas}\frac{\DD P}{\DD z}.
\end{equation}
In the above equation $\Phitot$ is the total gravitational potential (stellar disc, $\HI$ disc and dark matter), $P$ is the thermal pressure of the gas, assumed to be isothermal. For any chosen $\Phitot$, $\rhogas$ is constrained requiring
\begin{equation}
 \Sigmagas(R) = \int_{-\infty}^{+\infty} \rhogas(R,z)\dd z,
\end{equation}
where $\Sigmagas$ is an exponential disc model as in equation (\ref{for:expgas}), expressed in terms of surface density. The total potential is determined iteratively, following the scheme of \cite{Springel2005}. 

\subsubsection{Velocity distributions}

For simplicity, only for the dark-matter halo, we assume $\Phitot=\Phidm$ (i.e. the dark-matter potential) and we draw the halo phase-space positions directly from the Hernquist analytic isotropic DF. 

We assume that the stellar disc DF depends only on the energy $E$ and the third component of the angular momentum $\Lz$. Hence, the only non-vanishing moments of the stellar disc DF are $\hvRii = \hvzii \equiv \sigma^2$, $\hvphiii$ and $\hvphi$. We compute $\sigma$ and $\hvphiii$ from the Jeans equations, and, to sample the radial and vertical components of the stellar particle velocities, we assume that the velocity distributions in these two directions are Gaussian, with dispersion equal to $\sigma$. To compute the streaming velocity $\hvphi$ we rely on the epicyclic approximation, so  
\begin{equation}
    \hvphi = \sqrt{\hvphiii - \frac{\hvRii}{\eta^2}},
\end{equation}
where 
\begin{equation}
    \eta^2 = \frac{4}{R}\frac{\DD\Phitot}{\DD R}\biggl(\frac{3}{R}\frac{\DD\Phitot}{\DD R} +\frac{\DD^2\Phitot}{\DD R^2}\biggr)^{-1}.
\end{equation}
The azimuthal component of the stellar disc particles are then sampled from a Guassian with $\hvphi$ as mean and the r.m.s. velocity $\sqrt{\hvphiii-\hvphi^2}$ as standard deviation. 

The gas velocity field is instead composed only by the azimuthal component $\vphigas$ which satisfies the stationary Euler equation
\begin{equation}
 \vphigas^2 = R\biggl(\frac{\DD\Phitot}{\DD R} + \frac{1}{\rhogas}\frac{\DD P}{\DD R}\biggr).
\end{equation}

\subsubsection{Model parameters and equilibrium configuration}
\label{sec:comp}

The reference model derived in Section~\ref{sec:modelscomp} (see Table~\ref{tab:param}) fixes almost all the degrees of freedom needed to set the NGC 5474-like model of our simulations. We further adopt a stellar scale height $\zstar=0.15\hstar$ \citep{Kregel2002,Oh2015} while the gas is mono-atomic, with $T/\mu=3400\K$ ($T$ the gas temperature and $\mu$ the gas mean molecular weight). As a reference, for a neutral, hydrogen gas, this corresponds to $T=3400\K$.

We require all particles to have the same mass $\mpart=5000\Msun$. With this choice, given 
\begin{equation}\begin{split}
 & \Mvir  = 3.77\times10^9\Msun,  \\ 
 & \Mgas  = 1.82\times10^9\Msun,  \\ 
 & \Mstar = 4.11\times10^8\Msun, 
\end{split}\end{equation}
it follows
\begin{equation}\begin{split}
 & \Nh   = 708750, \\
 & \Ngas = 364000, \\
 & \Nstar = 82200, 
\end{split}\end{equation}
where $\Nh$, $\Ngas$ and $\Nstar$ indicate, respectively, the number of particles of the halo, the $\HI$ disc and the stellar disc\footnote{The dark halo of NGC 5474 is not sampled at radii larger than $3\rvir$, so the total mass represented with particles is $\simeq 0.94\Mvir$.}.

Since the dark-matter halo particles of NGC 5474 have been sampled as if the halo were in isolation (i.e. not accounting for the contribution of the discs to the total gravitational potential), the stellar disc was built in Maxwellian approximation and the discs provide a non-negligible contribution to the total gravitational potential, we expect all the models components to be close to equilibrium, but not exactly in equilibrium. To check how they respond to the presence of each other, and to let them shift towards an equilibrium state, we first run a simulation where NGC 5474 is evolved in isolation. The main features and the results of this simulation are described in Appendix~\ref{sec:DHeq}. In all the following hydrodynamic $N$-body models, we will take as ICs of NGC 5474 the output of the simulation of Appendix~\ref{sec:DHeq} after $0.98\Gyr$.

\begin{table*}
    \centering
    \caption{Main input parameters of the set of simulations of Section~\ref{sec:sim1}. From the left-hand to the right-hand column: name of the model (model's name); PB effective radius ($\Reff$); PB total stellar mass ($\Mb$); PB number of particles ($\Nb$); softening used for the PB particles ($\lPB$). The softening is computed requiring that the maximum force between the PB's particles should not be larger than the PB's mean-field strength (\citealt{Dehnen2011}). We notice that all the models components (halo, stellar disc, gas disc, PB) have different softenings. The PB particles have mass $\mpart=1667\Msun$. In each simulation the initial position of the PB centre of mass is at $\RPB=7\kpc$, with initial streaming velocity $\vc(\RPB)=42\kms$. The ICs of NGC 5474 correspond to the configuration of Appendix~\ref{sec:DHeq} taken after $0.98\Gyr$ (for further details see Tables~\ref{tab:param} and \ref{tab:siminput}).}\label{tab:simimput1}
    \begin{tabular}{lcccc}
    \hline\hline
    Models' name    &   $\Reff$ [$\kpc$]    &   $\Mb$ [$10^8\Msun$] &   $\Nb$   &   $\lPB$ [$\kpc$] \\
    \hline\hline    
    PB\_Re484\_M1.5 &       0.484       &   1.5     &       90000  &    0.029   \\
    PB\_Re484\_M1   &       0.484       &   1       &       60000  &    0.033   \\
    PB\_Re484\_M0.5 &       0.484       &   0.5     &       30000  &    0.042   \\
    PB\_Re320\_M1.5 &       0.320       &   1.5     &       90000  &    0.019   \\
    PB\_Re320\_M1   &       0.320       &   1       &       60000  &    0.022   \\
    PB\_Re320\_M0.5 &       0.320       &   0.5     &       30000  &    0.028   \\
    PB\_Re161\_M1.5 &       0.161       &   1.5     &       90000  &    0.0096  \\
    PB\_Re161\_M1   &       0.161       &   1       &       60000  &    0.011   \\
    PB\_Re161\_M0.5 &       0.161       &   0.5     &       30000  &    0.014   \\
    \hline\hline
    \end{tabular}
\end{table*}

\subsection{Realization of the PB}
\label{sec:PBICs}
The phase-space positions of the PB stellar and dark-matter particles are drawn directly from the components ergodic DFs. Starting from equation (\ref{for:ser1}), through an Abel inversion we retrieve the intrinsic density distribution $\rhostarPB$ of the stellar component. We complete the stellar and dark-matter density-potential pairs ($\rhostarPB,\PhistarPB$) and ($\rhodmPB,\PhidmPB$) solving the Poisson equation $\nabla^2\PhiastPB=4\pi G\rhoastPB$, where $*=(\star,\dm)$.

By means of an Eddington inversion \citep{BinneyTremaine2008} we compute numerically the ergodic DFs of stars ($\fstarPB$) and dark matter ($\fdmPB$) as
\begin{equation}
\fastPB(E) = \frac{1}{\sqrt{8}\pi^2}\frac{\dd}{\dd E}\int_0^E \frac{\dd \PhiPB}{\sqrt{E-\PhiPB}}\frac{\dd \rhoastPB}{\dd \PhiPB},
\end{equation}
where $*\in\{\star,{\rm dm}\}$, while $\PhiPB$ is the total potential $\PhiPB=\PhistarPB+\PhidmPB$. In case the PB is only made by stars, $\PhiPB=\PhistarPB$ and $*=\star$. 

Since in our simulations we will consider PBs with and without dark matter, and with stellar components ranging over different sizes and masses, for clarity, we will separately list in Sections \ref{sec:sim1pb} and \ref{sec:sim2pb} the PB parameters and the number of particles used in each simulation.

\section{Results}
\label{sec:sims}

\subsection{First hypothesis: the PB within the discs' plane}
\label{sec:sim1}

\subsubsection{Setting the PB parameters}
\label{sec:sim1pb}

\begin{figure*}
    \centering
    \includegraphics[width=.9\hsize]{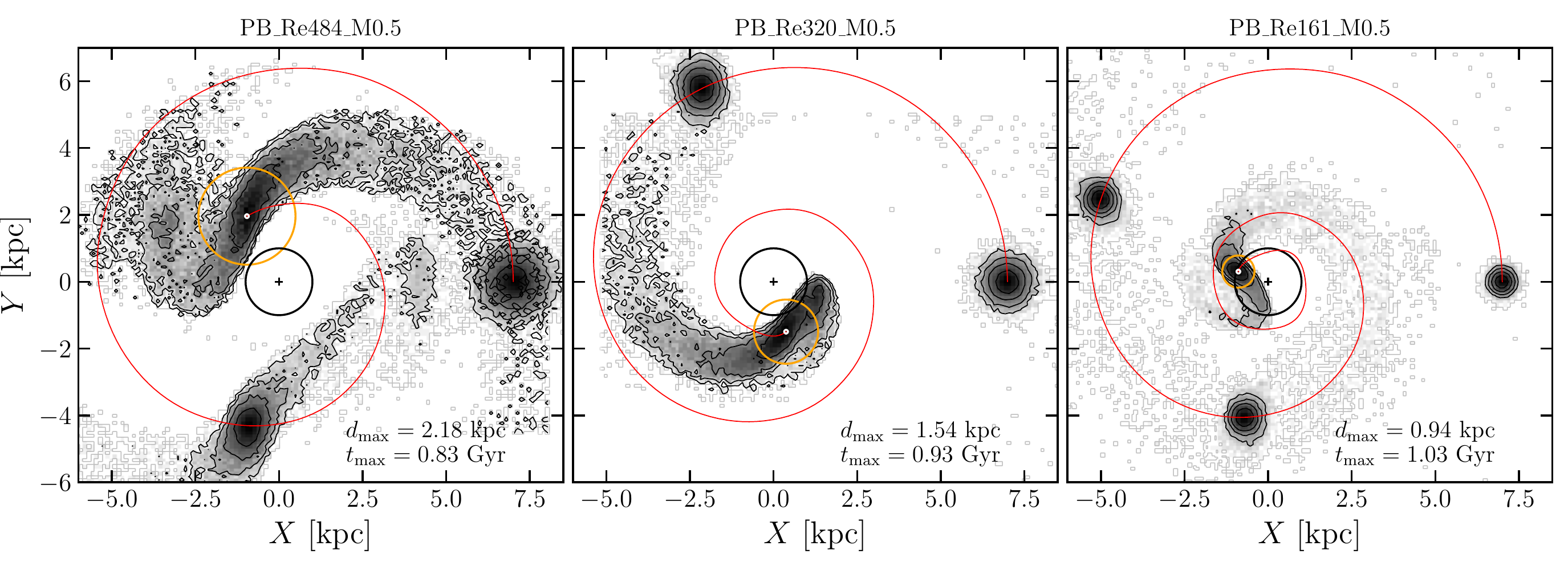}
    \includegraphics[width=.9\hsize]{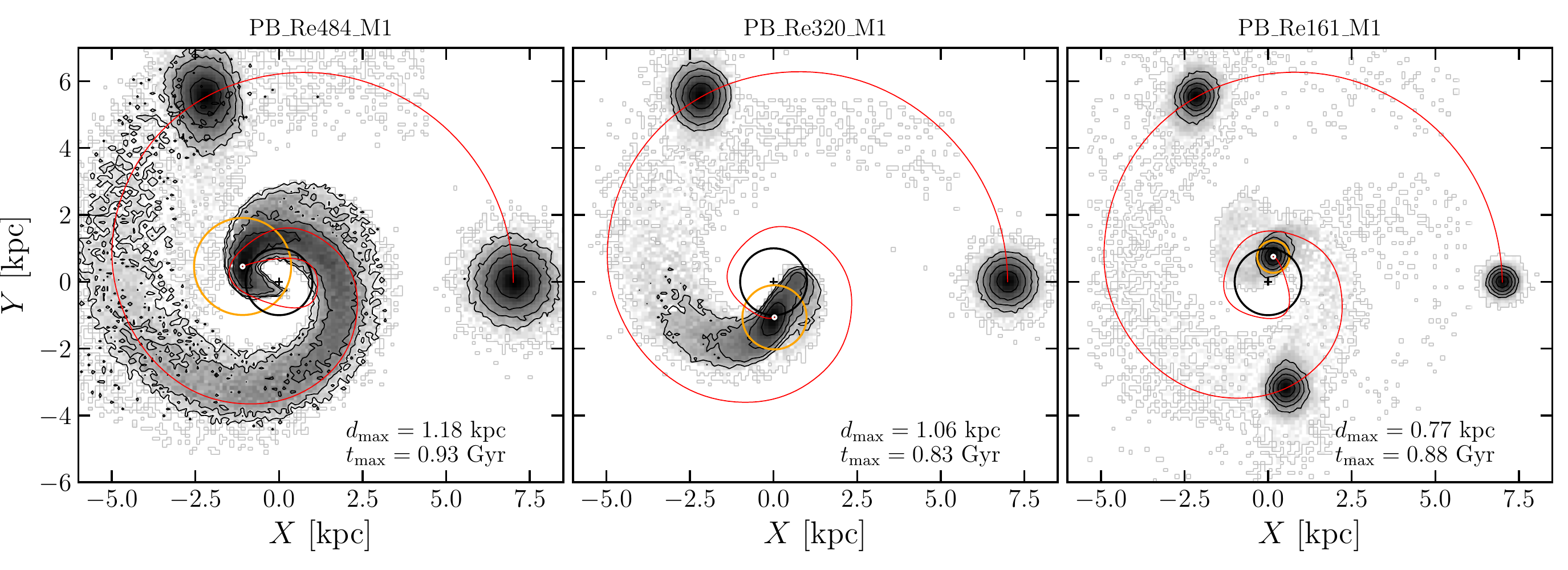}
    \includegraphics[width=.9\hsize]{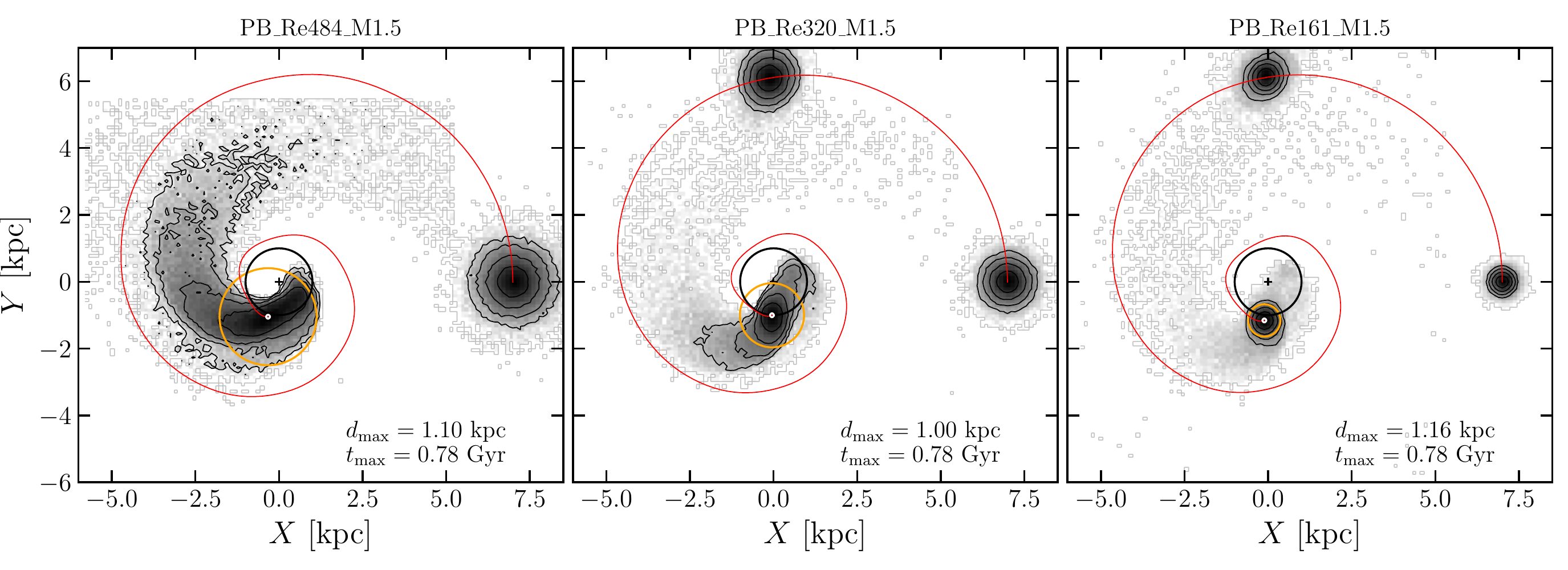}
    \caption{Trajectories of the centre of mass of the PBs (red curves) in all the hydrodynamical $N$-body models considered in Section~\ref{sec:sim1}. The top, middle and bottom rows of panels refer to models whose PB has an initial stellar mass $\Mb=0.5\times10^8\Msun$, $\Mb=10^8\Msun$ and  $1.5\times\Mb=10^8\Msun$, respectively. The initial PB effective radius decreases from the left to the right column of panels. In each panel we also show the PB spatial density distribution taken at few representative snapshots along its orbit, projected along the symmetry axis (so the $(X,Y)$-plane is the equatorial plane). In most cases, the trajectory is drawn until the PB reaches a distance of $\sim1\kpc$ from the centre (black circle). The PB centre is determined using the shrinking sphere method (\citealt{Power2003}).  Details on simulation parameters in Tables ~\ref{tab:param}, \ref{tab:simimput1} and \ref{tab:siminput}.}
    \label{fig:evolvedPB}
\end{figure*}

In our first set of simulations we explore cases in which the PB moves within the galaxy discs' plane, and we check whether the off-set can be reproduced as an outcome of the simulations keeping the shape of the PB smooth and regular and the kinematics of the galaxy's gaseous component unperturbed as observed.

Starting from the conclusions of Section~\ref{sec:model}:
\begin{itemize}
    \item we consider a PB made only of stars, without dark-matter halo;
    \item the PB centre of mass is at an initial distance $\RPB=7\kpc$ from the galaxy centre, on a circular orbit with an initial azimuthal velocity $\vc(\RPB)$, co-rotating with the $\HI$ and stellar discs. We expect the PB orbit to shrink because of dynamical friction and thus to reach $\RPB\approx 1\kpc$; 
    \item we focus on a PB with total initial stellar mass $\Mb=0.5\times10^8\Msun$, $\Mb=10^8\Msun$ and $\Mb=1.5\times10^8\Msun$. For each mass, we consider PBs with initial $\Reff=0.484\kpc$, $\Reff=0.320\kpc$ and $\Reff=0.161\kpc$. The case of $\Mb=1.5\times10^8\Msun$ is intended to account for the fact that, due to mass loss, the PB can reach $1\kpc$ with less than the upper limit of $10^8\Msun$ that we estimated in Section~\ref{sec:model}.
\end{itemize}
While we have required that the particles of all the components of NGC 5474 must have $\mpart=5000\Msun$, we relax this condition on the PB and we set its particles to be three times less massive. This allows us to sample the PB's phase space with a sufficiently large number of particles and to avoid an overwhelmingly high number of particles per simulation, which would just be computationally expensive with no particular gain in terms of accuracy. The PBs with $\Mb=0.5\times10^8\Msun$, $\Mb=10^8\Msun$ and $\Mb=1.5\times10^8\Msun$ are sampled with $\Nb=30000$, $\Nb=60000$ and $\Nb=90000$ particles, respectively, following the scheme of Section~\ref{sec:PBICs}. When these PBs are evolved in isolation for $10\Gyr$ they keep their equilibrium configuration.

We expect the PB to sink towards the system centre due to dynamical friction on a timescale $\tdf$, which we estimate as \citep{BinneyTremaine2008}
\begin{equation}\label{for:tdf}
    \tdf=\frac{1.17}{\ln\Lambda}\frac{\Mtot(\RPB)}{\Mb}\tc,
\end{equation}
where $\RPB$ is the PB distance from the centre, $\ln\Lambda$ is the Coulomb logarithm, $\tc\equiv2\RPB/\vc(\RPB)$ is the crossing time at $\RPB$ and $\Mtot$ is the total mass enclosed within $\RPB$. At $\RPB=7\kpc$, with $\vc(\RPB)\simeq42\kms$, we get
\begin{equation}\begin{split}\label{for:tdf2}
    & \tdf = 1.4-3.5\Gyr \text{\quad when\quad} \Mb=0.5\times10^8\Msun, \\
    & \tdf = 0.7-1.67\Gyr\text{\quad when\quad} \Mb=10^8\Msun, \\
    & \tdf = 0.43-1.15\Gyr\text{\quad when\quad} \Mb=1.5\times10^8\Msun, \\
\end{split}\end{equation}
The lower and upper limits over $\tdf$ are obtained assuming the typical values $\ln\Lambda\sim15$ and $\ln\Lambda\sim6$, respectively. According to these estimates we may expect the PB to decay towards the center of NGC 5474 on a very short timescale.

We run a total of 9 simulations and, in each of them, the ICs of NGC 5474 correspond to the equilibrium simulation of Appendix~\ref{sec:DHeq} after $0.98\Gyr$. After $0.98\Gyr$ the center of mass of NGC 5474 is in the origin of the system reference frame. Details on simulations and NGC 5474 parameters (e.g. softenings, number of particles) are listed in Tables~\ref{tab:param} and ~\ref{tab:siminput}. We will refer to these simulations as PB\_Re$X$\_M$Y$, where $X=484,320,161$ indicates the PB effective radius in pc, and $Y=0.5,1,1.5$ is the PB mass in units of $10^8$ (see also Table~\ref{tab:simimput1}). The simulations run for $4.1\Gyr$ and we use an adaptive timestep refinement with typical timestep values $\simeq0.5\Myr$.

\subsubsection{Results}
\label{sec:res}

\begin{figure}
    \centering
    \includegraphics[width=1\hsize]{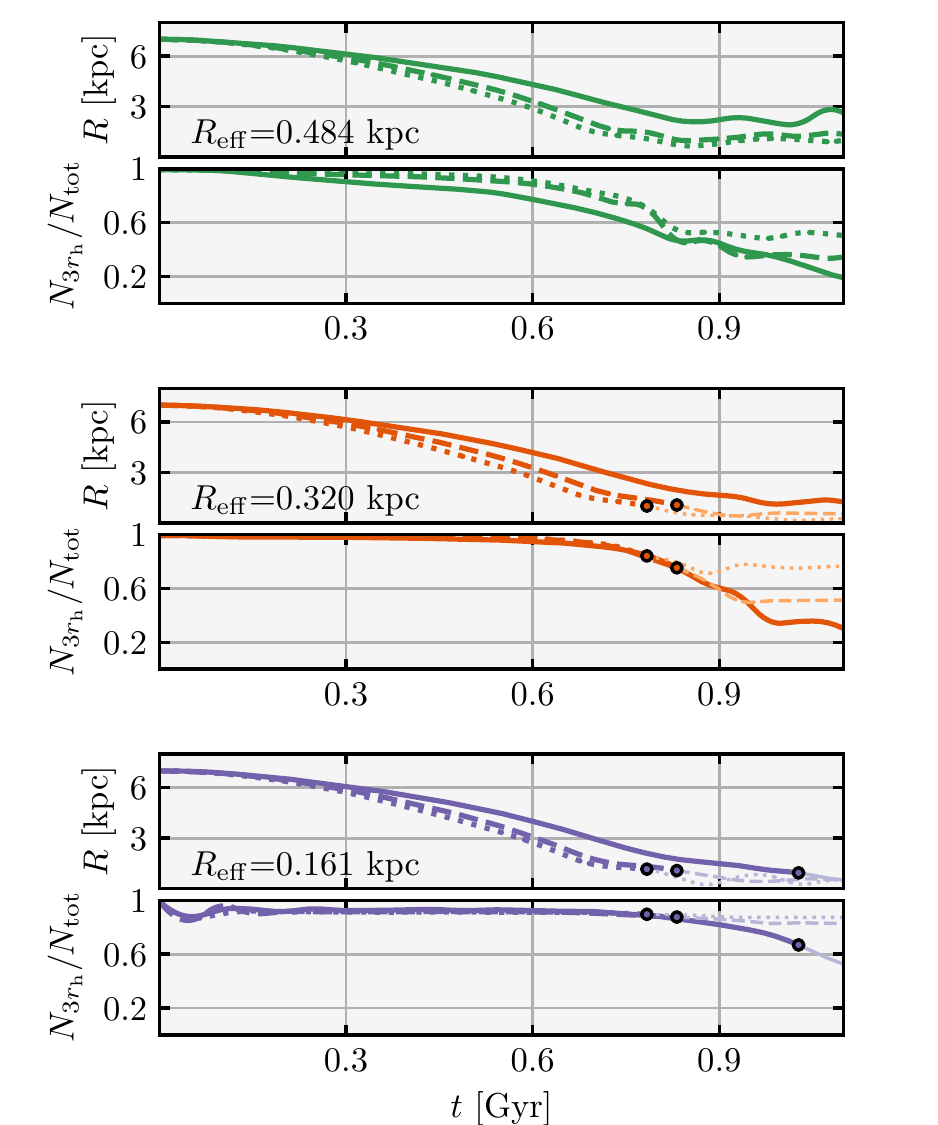}
    \caption{Top two panels: distance between the centre of mass of the PB and the galaxy's centre as a function of time (top) and fraction of mass enclosed within $3\rh$ as a function of time (bottom), with $\rh$ the PB half-mass radius as in the ICs. The green curves refer to PBs with initial $\Reff=0.484\kpc$. Middle two panels: same as the top panels, but for PB with initial $\Reff=0.320\kpc$ (orange curves). Bottom two panels: same as the top panels, but for PB with initial $\Reff=0.161\kpc$ (purple curves). The solid, dashed and dotted curves refer, respectively, to the PBs with initial mass $\Mb=0.5\times10^8\Msun$, $\Mb=10^8\Msun$ and $1.5\times\Mb=10^8\Msun$. The curves lighten when the PBs have reached a distance from the system's centre of $\approx 1\kpc$ (colored circles).}\label{fig:orbitsevo}
\end{figure}

Figure~\ref{fig:evolvedPB} shows the trajectories of the PBs in each of the nine simulations. Each orbit (red curve) is obtained connecting the centre of mass of the PB from consecutive snapshots and, alongside the orbit, each panel also shows the projected density distribution of the PB taken at few representative snapshots. The $(X,Y)$-plane is the plane of the orbit, co-planar with the discs' plane. We point out that in Fig.~\ref{fig:evolvedPB} we have assumed the galaxy's symmetry axis as line of sight, but, as long as the PB evolves on the discs plane, the term $\cos i$ provides a negligible correction for $i=21^{\circ}$, and we can anyway project the galaxy in such a way to align the off-set with one of the galaxy's principal axes.

As expected, the PBs with the largest initial $\Reff$ are distorted the most by the galactic tidal field. At $R\simeq2\kpc$, the PB of model PB\_Re484\_M0.5 has: i) lost approximately 60\% of its mass; ii) lost its spherical symmetry in favour of the formation of an elongated structure that has just started to wrap the galaxy centre; iii) developed extended and massive tidal debris, formed from the very beginning of the simulations. We find a similar outcome also in models PB\_Re484\_M1 and PB\_Re484\_M1.5, notwithstanding the higher PB mass which should make, in principle, the PB more resistant against the galaxy tidal force field. To estimate the mass loss we consider as particles belonging to the PB those that remain within 3$\rh$ ($\rh$ is the PB stellar half-mass radius in the ICs). As anticipated in Section~\ref{sec:model}, and looking at Fig.~\ref{fig:rt}, this result is not surprising given that the PB tidal radius is less than its initial half-mass radius at best.

\begin{figure*}
    \centering
    \includegraphics[width=1\hsize]{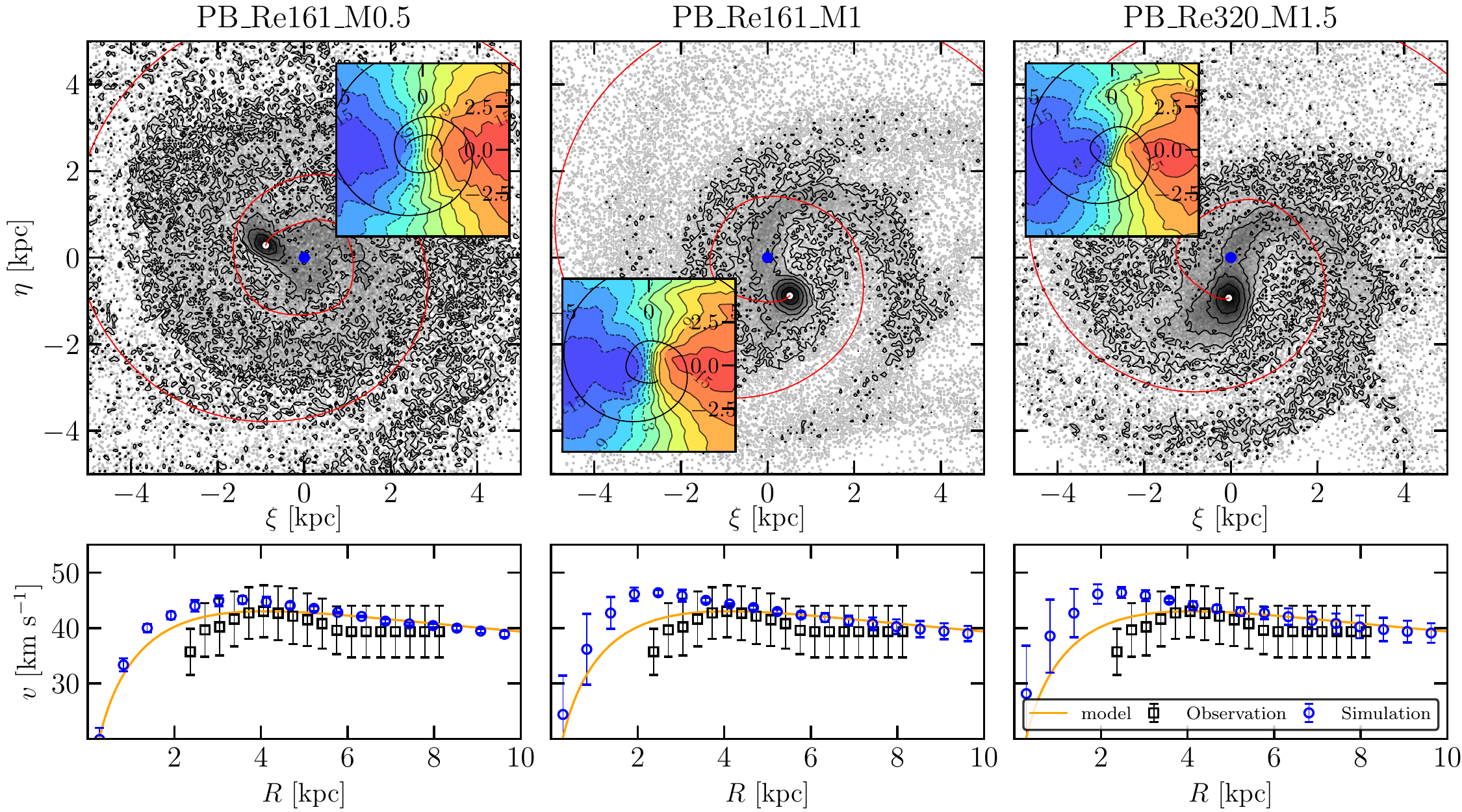}
    \caption{Top left panel: total (disc and PB) stellar projected density map computed from the configuration corresponding to the orbit's end-point of the hydrodynamical $N$-body model PB\_Re0.161\_M0.5 as in Fig.~\ref{fig:evolvedPB}. The system has been projected assuming an inclination of $i=21^{\circ}$, as in \citetalias{Rownd1994}. We show intensity contours equal to $\Smax/2^n$ with $\Smax$ the map's densest peak and $n=1,...,6$. The blue and white dots show, respectively, the discs kinematic centre and the PB centre and the full projected orbit is shown with a red line. We notice that the disc kinematic center also corresponds to the centre of the $(\xi,\eta)$ plane. The small inset shows the $\HI$ line-of-sight velocity map, derived in the same portion as in the main panel. The $\HI$ velocity map has been obtained as in \citetalias{Rownd1994}, binning with pixels $0.33\kpc\times0.33\kpc$ wide, once we have assumed the distance $d=6.98\Mpc$, and with velocity contours separated by $3\kms$. The approaching arm is shown with red colors and solid black curves, while the receding arm with blue colors and dashed-black curves. Bottom left panel: $\HI$ circular speed as a function of the distance from the galaxy kinematic center (blue points with error bars) from the same snapshot as in the top left panel, compared to the galaxy’s deprojected rotation velocity curve computed in Section~\ref{sec:hd} (black points with error bars). The orange curve shows the circular speed of the analytic model of NGC 5474. Middle panels: same as the left panels but for the $N$-body model PB\_Re0.161\_M1. Right panels: same as the left panels but for the $N$-body model PB\_Re0.320\_M1.5. Details on simulation parameters in Tables ~\ref{tab:param}, \ref{tab:simimput1} and \ref{tab:siminput}.}
    \label{fig:stellarzooms}
\end{figure*}


\begin{figure*}
 \begin{subfigure}[b]{.325\textwidth}
 \includegraphics[width=1\hsize]{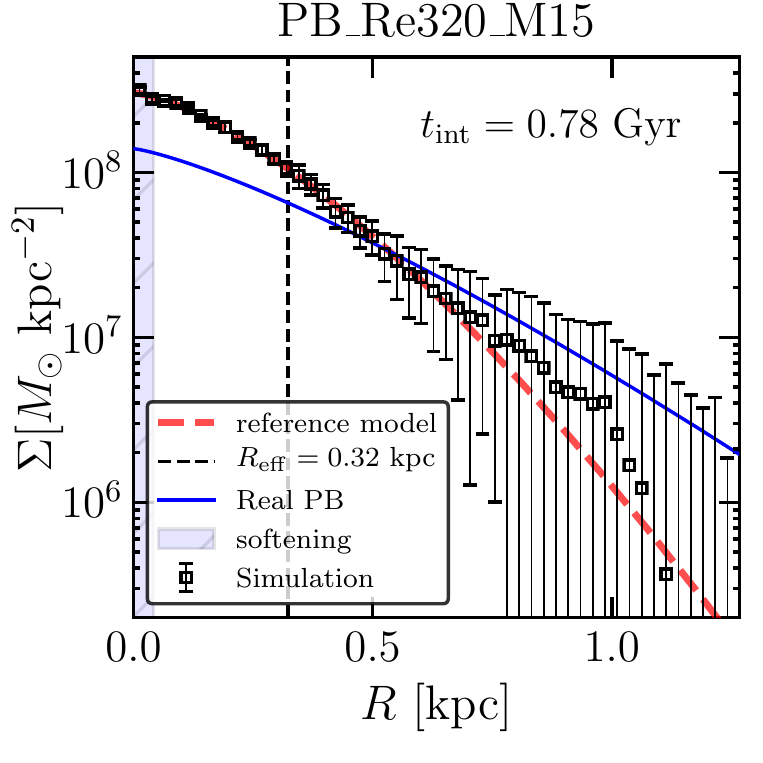}
 \end{subfigure}\quad
 \begin{subfigure}[b]{.575\textwidth}
  \includegraphics[width=1\hsize]{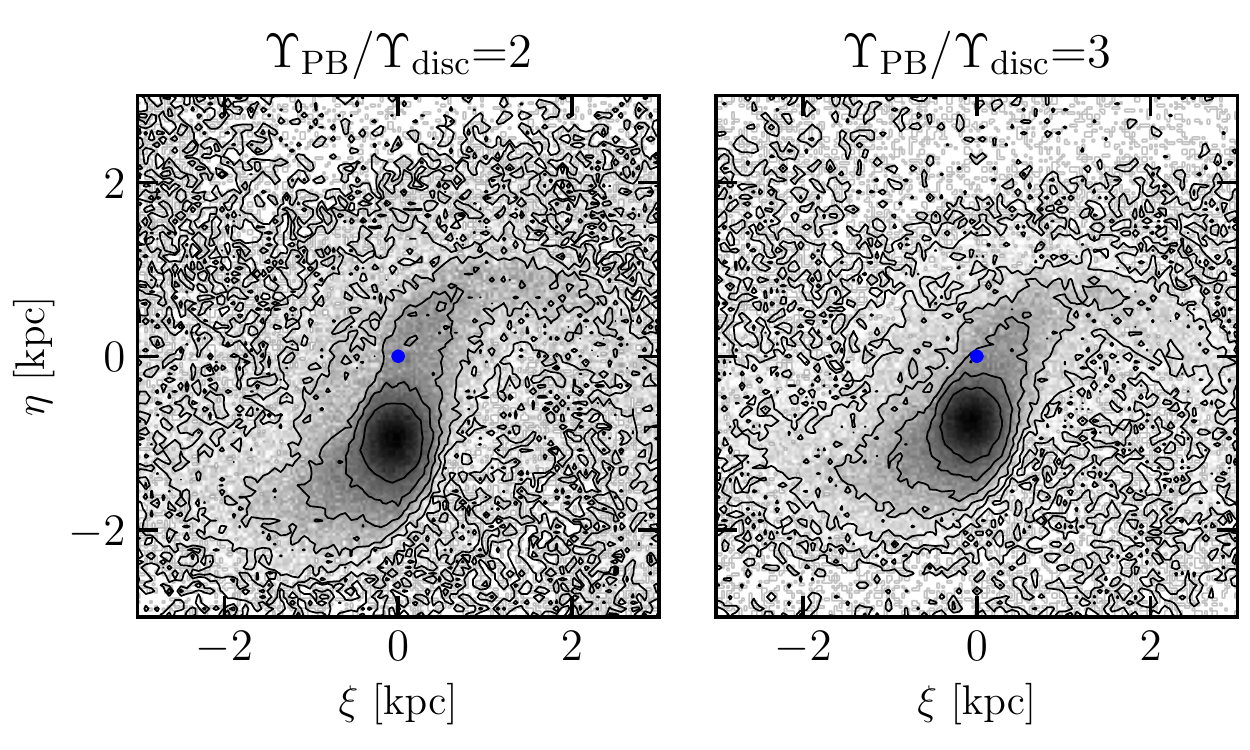}
 \end{subfigure}
 \caption{Left panel: projected density profile of the PB computed from the density map of the right panel of Fig.~\ref{fig:stellarzooms} (black points with error bars), superimposed to the reference best fitting model, obtained as in Section~\ref{sec:res} (red dashed curve). As a comparison, we show the \Sersic model resulting from the fit of \citetalias{Bellazzini2020} (blue curve), so $m=0.79$, $\Reff=0.484\kpc$, while we have imposed the same total mass as the best model ($\Mb\simeq10^8\Msun$, see Table~\ref{tab:pbsim}). The vertical black-dashed line marks the effective radius of the reference model. Middle panel: total stellar (disc and PB) surface brightness map as in the right hand panel of Fig.~\ref{fig:stellarzooms}, where we have assumed a different mass-to-light-ratio ($\Upsilon$) for stars belonging to the stellar disc and the PB. The PB is twice as luminous as the stellar disc. The blue dot, also the centre of the $(\xi,\eta)$ plane, shows the kinematic centre of the stellar and $\HI$ discs. Right panel: same as the middle panel, but the PB particles are three times as luminous as the stellar disc ones.}\label{fig:PBpost}
\end{figure*}

For each simulation, Fig.~\ref{fig:orbitsevo} shows the projected distance between the PB centre of mass and the galaxy centre, and the PB bound stellar mass fraction as a function of time. The systems are projected as in Fig.~\ref{fig:evolvedPB}, assuming as line of sight the symmetry axis, so $R\equiv\sqrt{X^2+Y^2}$. The main driver of the PB evolution is the dynamical friction: in a very short time-scale (less than $0.8\Gyr$; see also equation~\ref{for:tdf2}) all the PBs reach $R\sim2\kpc$ and, as expected, the most massive PBs sink faster than the least massive ones (equation \ref{for:tdf}; top panel of Fig.~\ref{fig:orbitsevo}). Among the PBs with initial $\Reff=0.320\kpc$, those of models PB\_Re320\_M1 and PB\_Re320\_M1.5 reach $R=1\kpc$ losing only 20\% of their original mass, while that of model PB\_Re320\_M0.5 has experienced substantial mass loss already at $R\simeq1.54\kpc$. The most compact PBs (PB\_Re161\_M0.5, PB\_Re161\_M1 and PB\_Re161\_M1.5), instead, provide a great resistance against the galaxy tidal force field  ($\rt\simeq7\Reff$, see Fig.~\ref{fig:rt}), and reach the galaxy centre losing from 10\% to 30\% of their total initial mass, without developing significant tidal tails.

Based on the aforementioned features, we exclude from any further analysis the PBs with initial $\Reff=0.484\kpc$ and the one from model PB\_Re320\_M0.5, which immediately depart from equilibrium and are severely distorted by the strong tidal force field of NGC 5474. We notice that when the initial mass is $\Mb=1.5\times10^8\Msun$, in some cases the PB reaches $1\kpc$ with less than $10^8\Msun$ (the mass upper limit estimated in Section~\ref{sec:model}). 
For this reason, we ran additional simulations with a PB initial mass as high as $\Mb=2\times10^8\Msun$. Even though these configurations reach the galactocentric $1\kpc$ distance with a bound mass of $10^8\Msun$ or more, when $\Reff=0.484\kpc$ the PB still develops pronounced tidal features, when $\Reff=0.320\kpc$ and $\Reff=0.161\kpc$ it presents the same structural properties 
of its $1.5\times10^8\Msun$ analogs, and they are not shown here for the sake of synthesis.

To make any comparison between simulations and observations coherent, in Fig.~\ref{fig:stellarzooms} we add the contribution due to the stellar disc of NGC 5474 to the PB projected density maps of some of the remaining models, taken when they are $1\kpc$ away from the galaxy centre (i.e. the orbits end-point of Fig.~\ref{fig:evolvedPB}). As prototypical cases, we selected models PB\_Re161\_M0.5, PB\_Re161\_M1 and PB\_Re320\_M1.5. The greyscale extends over two order of magnitudes from the highest density peak, which would be compatible with a difference of five magnitudes from the map brightest point if we assume the same mass-to-light ratio for all the particles. In this case, the systems have been projected assuming $i=21^{\circ}$, and we have called the plane of the sky the $(\xi,\eta)$-plane, with $\xi\equiv X$. For instance, such an image displays patterns that are somehow reminiscent of the galaxy stellar map of Fig.~\ref{fig:ngc5474} or Fig. 5 by \citetalias{Bellazzini2020}, both obtained through the LEGUS \citep{Calzetti2015} photometric catalogue based on HST ACS images. In all cases, the perturbation provided by the PB induces the formation of a loose spiral structure in a region $2-3\kpc$ around the discs' centre. The spiral arms are similar to the observed pattern of NGC 5474 \citepalias{Bellazzini2020}, they are not necessarily symmetric with respect to the discs' centre and also the gaseous component develops a spiral structure that follows the optical stellar disc \citepalias{Rownd1994}. In the small insets of Fig.~\ref{fig:stellarzooms} we show the corresponding $\HI$ line-of-sight velocity field maps as in the main panels. The velocity maps are derived using the same spatial and velocity resolutions as the one from the \citetalias{Rownd1994} map: pixels $0.33\kpc\times 0.33\kpc$ wide and velocity contours separated by $3\kms$. The solid black lines and redder colors mark the disc's approaching arm, while the black dashed curves and bluer colors indicate the receding arm. Once the resolution has been downgraded to the same one of the observations, almost all the $\HI$ velocity maps appear smooth and regular, and consistent with \citetalias{Rownd1994}. A different perspective is given by the $\HI$ rotation curves shown in the bottom panels of Fig.~\ref{fig:stellarzooms}. Most of the differences that are clearly visible between the rotation curves from the simulations and the analytic model are confined within $2-3\kpc$, where we barely have kinematic information. However, as predicted in Section~\ref{sec:model}, the rotation curves start to show appreciable differences with respect to the de-projected rotation curve we derived when the PB mass is close to $\simeq10^8\Msun$.

We rederived some of the structural parameters of the PBs from Fig.~\ref{fig:stellarzooms} fitting their projected density distribution with a \Sersic model, as in \cite{Fisher2010} and \citetalias{Bellazzini2020}, and computing their average axis-ratio $q$ through elliptical isodensity contours. We measure the axis ratio $q\equiv c/a$, with $c$ and $a$ the semi-minor and semi-major axes, respectively \citep[see][]{Lau2012}, computing the inertia tensor of the projected densities maps in an ellipse with semi-major axis $2\Reff$, centred on the map densest point. To produce the projected density distribution: i) we evaluate the PB centre on the plane of the sky as the densest/brightest point; ii) we bin the particles in 50 circular annuli, linearly equally spaced, and extending out to $4\Reff$, with $\Reff$ as in the models' ICs; iii) we subtracted the stellar disc contribution, estimated from a wider and more distant concentric circular annulus.

A \Sersic model is specified by the parameters $\{\Reff,m,\Mb\}$, as in equation (\ref{for:ser1}) of Section~\ref{sec:PB}, replacing surface brightness with stellar surface mass density, assuming constant mass-to-light ratio. The model's log-likelihood is 
\begin{equation}\label{for:chi2PB}
    \ln\LL = -\frac{1}{2}\sum_i\biggl(\frac{\Sigma(\Rsnapi) - \Sigsnapi}{\delta\Sigsnapi}\biggr)^2,
\end{equation}
where the points $\{\Rsnapi,\Sigsnapi,\delta\Sigsnapi\}$ are the PB projected density profile, and the sum extends over the total number or radial bins. The fit is performed using a Bayesian approach relying on MCMC, adopting the same scheme as in  Section~\ref{sec:method} and using flat priors over the model parameters. We list the parameters resulting from the fit in Table~\ref{tab:pbsim}.

On the plane of the sky, the PB of NGC 5474 appears round and regular ($q\simeq1$). In our analysis, the only PB that keeps its circular symmetry is in model PB\_Re161\_M1 ($q\simeq0.9$), while we estimate $q\simeq0.67$ and $q\simeq0.76$ for models PB\_Re161\_M0.5 and PB\_Re320\_M1.5, respectively. According to the \Sersic fit, the PBs with initial $\Reff=0.161\kpc$ are more extended, on average, by a factor of $\sim1.3$ with respect to the initial system, while the PB with initial $\Reff=0.320\kpc$ by a factor 1.15 (see Table~\ref{tab:pbsim}). As a reference, the left panel of Fig.~\ref{fig:PBpost} shows the projected density profile of the PB from model PB\_Re320\_M1.5, together with the best fitting \Sersic model. In this case, the large errorbars are caused by having imposed circular symmetry in the derivation of the profile, even though the system is not strictly circularly symmetric. 


\begin{table}
 \begin{center}
  \caption{Models' parameters as a result of the fit with the \Sersic model of Section~\ref{sec:res}. The columns, from left to the right, list: the model's name; the \Sersic effective radius ($\Reff$); the \Sersic index ($m$); the PB total mass ($\Mb$); the axis ratio ($q$). For the parameters of the \Sersic model, we take as measure of that parameter the $50^{\tth}$ percentile of its marginalized one-dimensional distribution, while the uncertatites are estimated from the 16$^{\tth}$, 50$^{\tth}$ and 84$^{\tth}$ percentiles of the corresponding marginalized one-dimensional distribution.}\label{tab:pbsim}
  \begin{tabular}{lccc}
   \hline\hline
    Model   &   PB\_Re161\_M0.5  &   PB\_Re161\_M1  &    PB\_Re320\_M1.5 \\
   \hline\hline
    $\Reff/\kpc$    &   $0.22\pm0.01$   &   $0.21\pm0.01$   &   $0.37\pm0.02$  \\
    $m$             &   $0.64^{+0.05}_{-0.04}$   &   $0.64\pm0.02$   &   $0.78\pm0.04$  \\
    $\logten\frac{\Mb}{\Msun}$&   $7.53\pm0.03$ & $7.93\pm0.01$ &   $8.14\pm0.2$        \\
    $q$             &   $0.67$          &   $0.90$          &   $0.76$                  \\
   \hline\hline
  \end{tabular}
 \end{center}
\end{table}

The case of model PB\_Re320\_M1.5 is probably the most intriguing one. Even if its PB is too flattened to look like the observed stellar system of NGC 5474 ($q\simeq0.76$), the fit with the \Sersic model provides $\Reff=0.37\pm0.02\kpc$, the closest to $\Reff=0.484\kpc$ among the cases explored. Since in Fig.~\ref{fig:stellarzooms} we have assumed the same mass-to-light ratio for the stellar disc and PB particles, it may be worth asking how would the PB look like if we made its particles more luminous. Assuming different mass-to-light-ratios $\UpsilonPB$ and $\Upsilondisc$ between stars of the PB and the galaxy's stellar disc, respectively, the middle and right panels of Fig.~\ref{fig:PBpost} show the surface brightness maps obtained from a zoom of the right panel of Fig.~\ref{fig:stellarzooms}. In the 3.6 $\mu$m band, if $\UpsilonPB$ ranges between 0.5 and 0.8, then $\UpsilonPB/\Upsilondisc=2$ (middle panel) implies a disc stellar population $\simeq1-3\Gyr$ old (see also Sections~\ref{sec:modelscomp} and \ref{sec:PB}). Although very extreme, we also examined the case with $\UpsilonPB/\Upsilondisc=3$ (right panel). The shades of colours are the same as in Fig.~\ref{fig:stellarzooms}. For these PBs we have rederived the axis ratios and the structural parameters resulting from a \Sersic fit, but we do not find any significant difference with respect to the case with uniform $\Upsilon$ in axis ratio, size and concentration.

To summarize, we find that a system with a large initial size steps away from its equilibrium state on a timescale so short to make very unlikely the possibility that it represents the observed PB of NGC 5474. It seems implausible that these systems may produce the observed off-set as the result of orbital decay as well as it seems implausible that the PB, if misplaced from the galaxy centre, would last long enough without developing visible debris or tidal tails. Furthermore, in this latter case, the problem of the mechanism that may have caused the off-set would still be an open question. The most compact PBs, at least over the range of masses explored, move towards the centre without producing detectable perturbation in the $\HI$ velocity field but, in spite of this, none of them reproduce the observed properties of the PB of NGC 5474. Not even the compromise of a massive and intermediate sized system (which can in principle resist the tidal force field, and expand to the required size during its evolution) is anyway close to the desired outcome. In fact, we find that a stellar component with these properties flattens to an extent inconsistent with the observations, independently on the luminosity of its stellar population. 

We do not expect an increase of the initial mass to improve the chances of survival of the PB: even though it would make it more resilient against the gravitational tidal field force, according to Section~\ref{sec:model}, the smoothness of the $\HI$ velocity field would start to be compromised. Instead, objects less massive than $0.5\times10^8\Msun$ would have a final mass (measured when the system is $1\kpc$ off-centred) less than the one we would expect on the basis of the luminosity of the PB stellar population, as discussed in Section~\ref{sec:PB}.

\begin{figure*}
    \centering
    \includegraphics[width=.95\hsize]{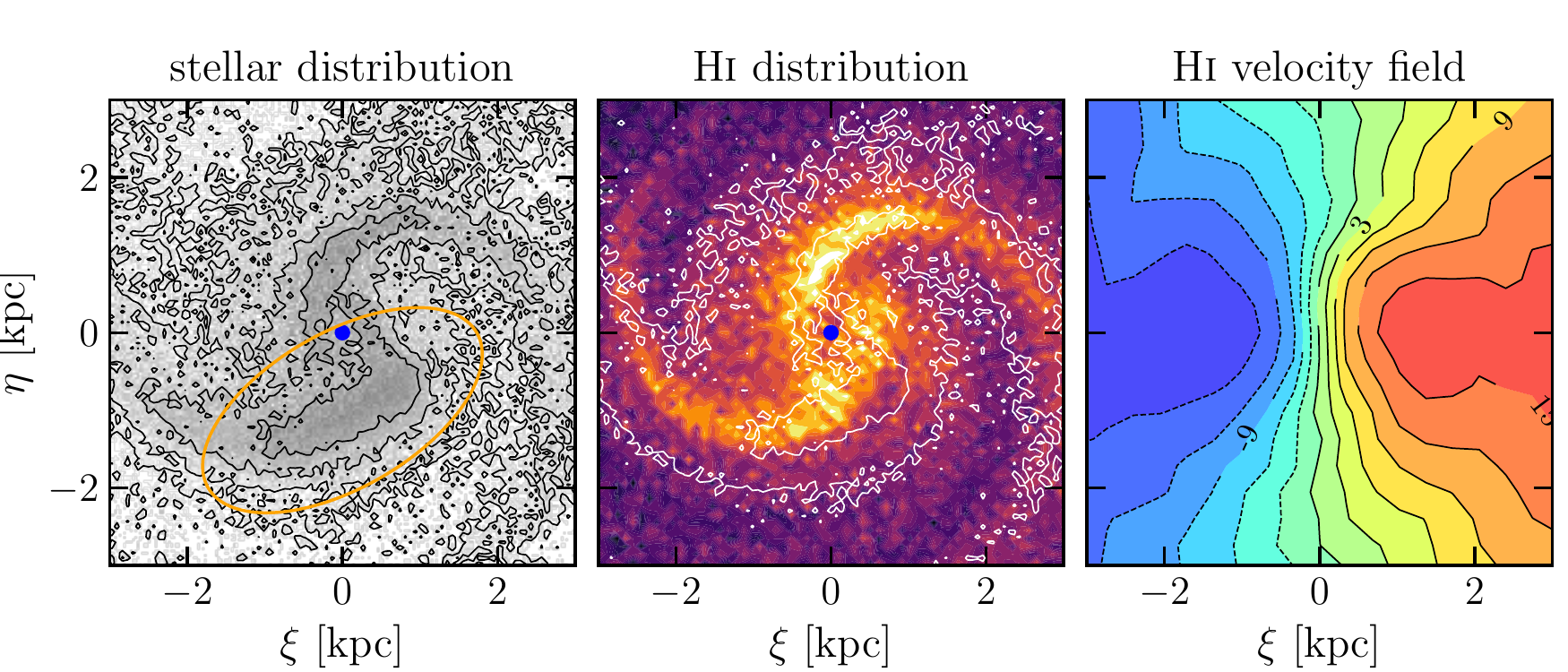}
    \caption{Left panel: stellar projected density map (stellar disc and PB) from the hydrodynamic $N$-body model PB\_Re484\_M1 taken after $\tint=0.83\Gyr$. The system has been projected as in Fig.~\ref{fig:stellarzooms}, assuming $i=21^{\circ}$, with $\xi\equiv X$. The centre of the $(\xi,\eta)$ plane is the kinematic centre of the gaseous disc (blue dot), while the intensity contours are at $\Smax/2^n$ with $\Smax$ the map's densest peak and $n=1,...,4$. The orange ellipse shows the position of the over-density. Middle panel: $\HI$ projected density map. The colour scale is such that yellow corresponds to high-density regions while purple to low-density regions. As a comparison, the white curves show the stellar isodensity contours as in the left panel. For clarity, we only show stellar isodensities $\Smax/2^n$ corresponding to $n=1,4$. Right panel: $\HI$ velocity field map. The solid curves and the redder colours mark the disc's approaching arm while the dashed curves and the bluer colours the disc's receding arms. The contours are separated by $3\kms$ and each pixel is $0.33\kpc\times0.33\kpc$ wide, as in \citetalias{Rownd1994}.}
    \label{fig:SW}
\end{figure*}

Even if we limited ourselves in studying only few specific orbits, as long as PB moves within the galaxy's equatorial plane, it seems likely that different orbits (for instance, more radial orbits, or with low inclination) would not behave so differently from the ones that we have considered given the dominant effects of dynamical friction and the strong tidal force field of NGC 5474.  


\subsubsection{The SW over-density}
\label{sec:sw}

The left panel of Fig.~\ref{fig:SW} shows the projected stellar density map from model PB\_Re484\_M1 after $\simeq1\Gyr$. We show a zoomed-in view of the discs' centre, the line of sight is inclined of $i=21^{\circ}$ with respect to the symmetry axis, and we have used the same scheme of colours as in Fig.~\ref{fig:stellarzooms}. We have shown that the PB of the hydrodynamical $N$-body model PB\_Re484\_M1 is disrupted by the tidal force field of NGC 5474 and perturbs the central stellar and gas distribution of NGC 5474 while sinking towards the centre. In the left panel of Fig.~\ref{fig:SW}, we mark with an orange ellipse what is left of it after $1\Gyr$, when its centroid is $\sim1\kpc$ away from the galaxy's centre.

We cannot help pointing out the similarities between this structure and the SW over-density of NGC 5474. We mentioned the SW over-density as one of the peculiarities of NGC 5474: it is a large substructure extending to the South-West of the PB of NGC 5474 (see Fig.~\ref{fig:ngc5474}), mostly dominated by old-intermediate age stars, whose structure is not associated with the overall spiral pattern \citepalias{Bellazzini2020}. When the PB of model PB\_Re484\_M1 is dismembered, its remnants shape into an elongated and wide structure appearing denser than the stellar disc's centre. The over-dense region partially brightens the galaxy's spiral structure with its tidal tails, and the spiral arms are also slightly traced by the $\HI$ distribution (middle panel). As noted by \citetalias{Bellazzini2020}, the SW over-density seems to be correlated to a local minimum of the $\HI$ distribution \citepalias{Kornreich2000}, which is partially consistent with Fig.~\ref{fig:SW}. We find a similar configuration also in model PB\_Re320\_M0.5. 

Reproducing the observed properties of the SW over-density of intermediate-old stars of NGC 5474 is beyond the scope of this work and it would require a systematic search of the wide parameter space (different orbits inclinations, eccentricity, initial velocity etc). However, as a by product, our simulations have produced configurations that are worth commenting as a viable channel for the formation of this substructure and, in general, of the substructures traced by old-intermediate age stars in the disc of NGC5474. An interaction with M 101 may not be enough to explain the SW over-density (\citetalias{Bellazzini2020}; \citealt{Mihos2013}) and we showed that the hypothesis that the SW over-density can be the remnant of a disrupted system, unrelated to the PB and M 101, as proposed, for instance, by \citetalias{Bellazzini2020}, is plausible and compatible with the smooth $\HI$ velocity field map (Fig~\ref{fig:SW}, right panel).

\subsection{Second hypothesis: the PB as a satellite}
\label{sec:sim2}
\subsubsection{Setting the PB and the halo}
\label{sec:sim2pb}

In this second set of simulations, we shall consider the scenario that would attribute the off-set as the apparent position of the PB in the disc to projection effects of an external system (an unbound galaxy or a bound satellite) crossing the line of sight (\citetalias{Rownd1994}; \citealt{Mihos2013}; \citetalias{Bellazzini2020}). By means of radial velocity measurements from emission (absorption) lines of stars from the disc (PB), \citetalias{Bellazzini2020} constrained the maximum line-of-sight velocity difference between the two to $\sim50\kms$, less than the circular speed of NGC 5474. This feature implies a similar distance and, together with the fact that disc and PB have stellar populations with comparable age, probably implies a common history as well. As such, we explore cases where the PB is an external satellite galaxy, moving around NGC 5474 onto motivated orbits, rather than an un-related foreground galaxy. We try to explore the possibility that some of the other peculiarities observed in NGC 5474 (the $\HI$ warp, \citealt{vanderHulst1979}; the SW over-density, \citetalias{Bellazzini2020}) may be caused by the gravitational interaction with such a satellite and, more generally, whether the observed properties of NGC 5474 are consistent with the presence of an orbiting satellite.



\begin{figure*}
    \centering
    \includegraphics[width=.49\hsize]{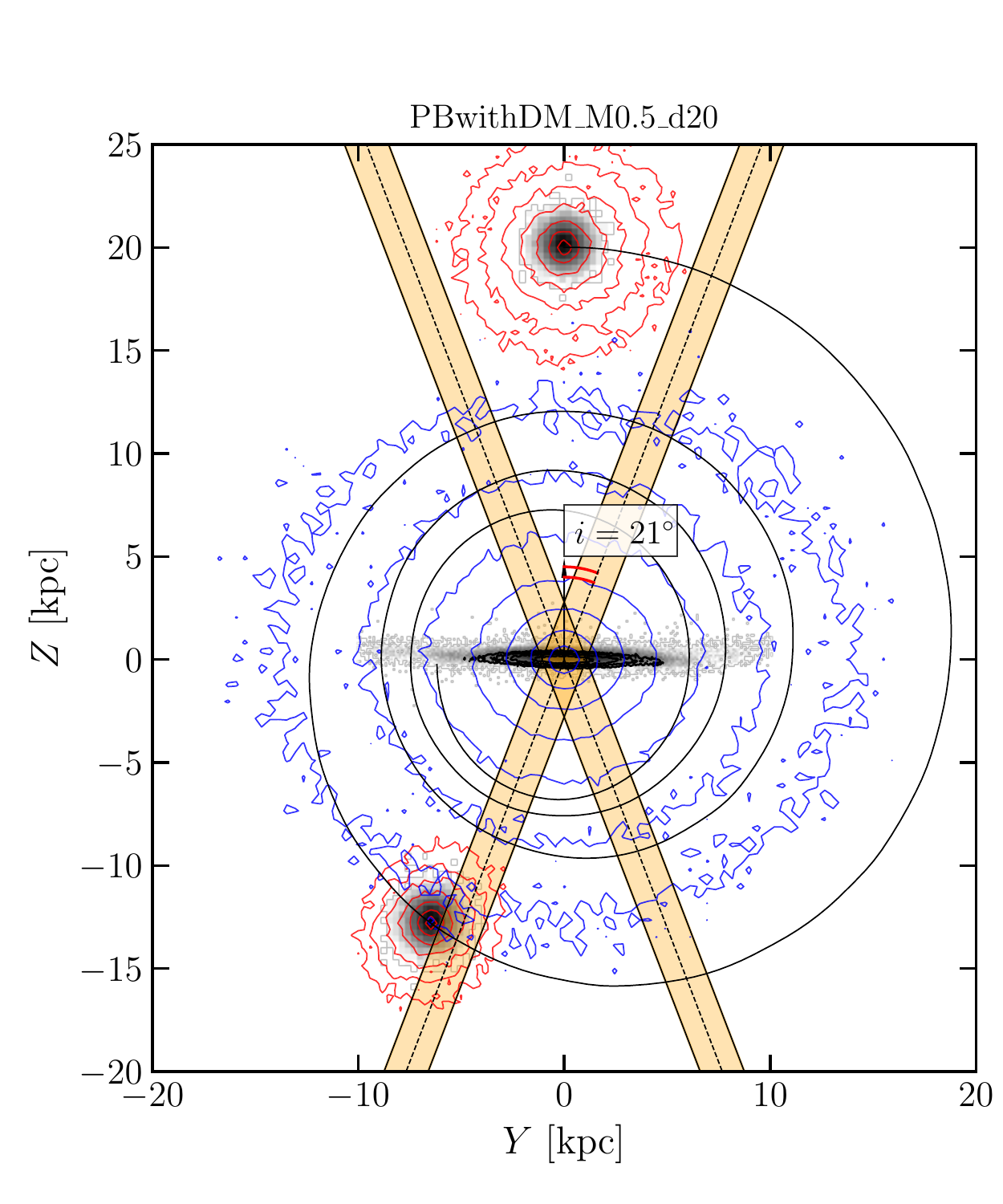}
    \includegraphics[width=.49\hsize]{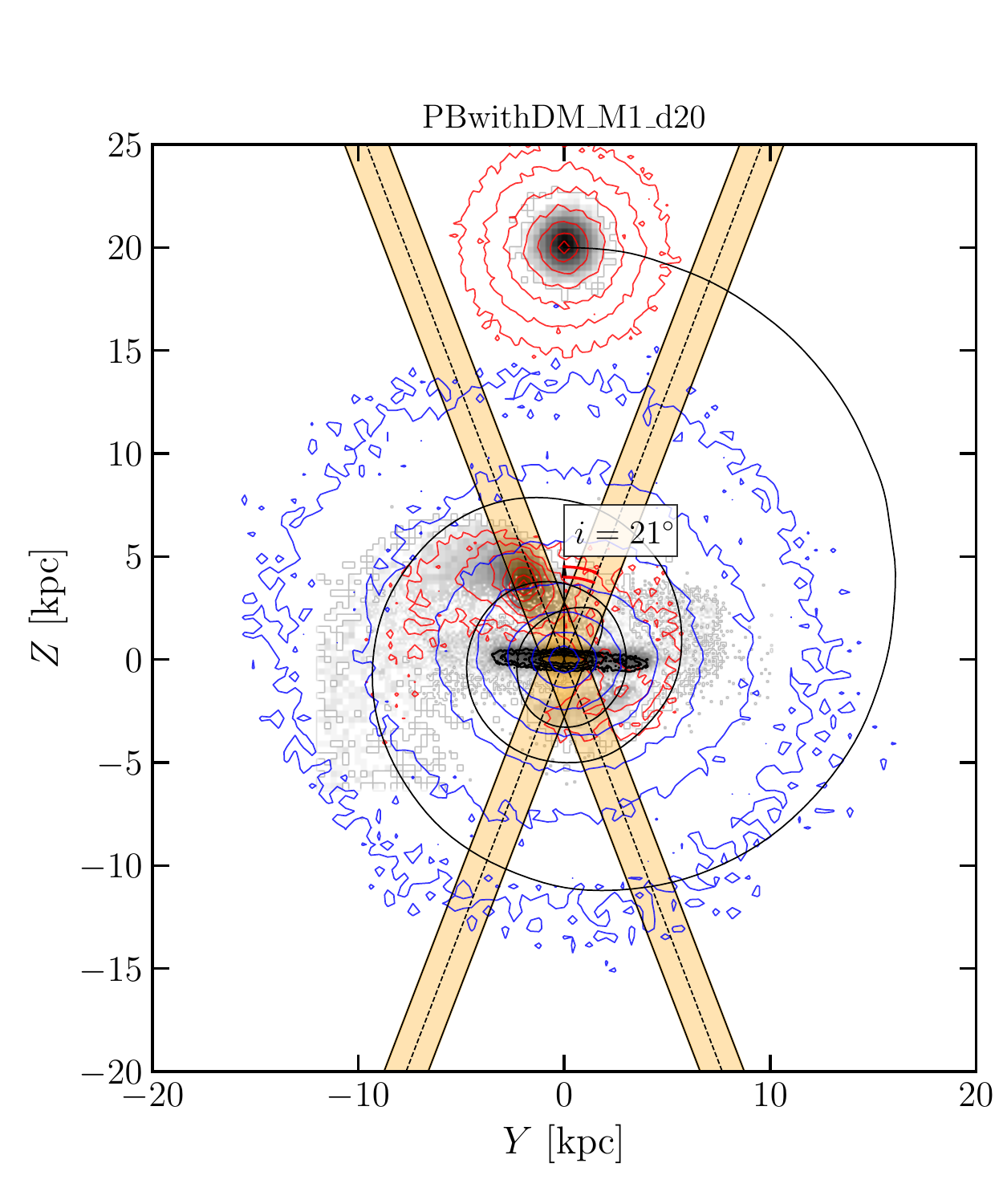}
    \caption{Left panel: trajectory around NGC 5474 (black curve) made by the PB of the hydrodynamical $N$-body model PBwithDM\_M0.5\_d20. The PB centre of mass starts from $(\xii,\yi,\zi) = (0,0,20\kpc)$, with initial velocity $\vc(\zi)=33\kms$ in the $y$-direction, in a Cartesian reference frame whose $(x,y)$-plane is the galaxy's disc plane. The orbit is shown for $\simeq6\Gyr$. The system has been projected assuming as line of sight the $x$-axis, so $Y\equiv y$ and $Z\equiv z$. The edges of the two orange bands show the different lines-of-sight ($i=21^{\circ}$) that would produce an off-set of $1\kpc$ on the plane of the discs, compatible with observations, when the PB centre crosses them. The stars of the PB are embedded with a dark-matter halo, whose isodensity contours are marked in red. We show two different PBs, one corresponding to the ICs and one corresponding to a crossing of one of the lines of sight. As a comparison, we also show the projected distributions of the stellar disc of  NGC 5474 (black points) and the NGC 5474 dark-matter halo isodensities (blue curves). Right panel: same as the left panel, but for the $N$-body model PBwithDM\_M1\_d20. In the latter case, the orbit is shown for $\simeq3\Gyr$.}
    \label{fig:orbitDM}
\end{figure*}

We focus on orbits that pass right above the galaxy's symmetry axis, starting from $(\xii,\yi,\zi)=(0,0,20\kpc)$, with $x,y,z$ the axes of a Cartesian reference frame whose $(x,y)$-plane is the galaxy's plane of the discs. The only non-zero component of the initial velocity of the PB centre of mass is in the $y$-direction, with modulus $\vc(\zi)=33\kms$. Since the PB moves far from the discs plane we relax the condition on its total mass and we embed it in a realistic dark-matter halo, following the scheme described in Section~\ref{sec:PB}. We focus on the two cases of a PB with a stellar mass of $\Mb=0.5\times10^8\Msun$ or $\Mb=10^8\Msun$. The PB halo parameters are chosen as in Section~\ref{sec:PB} and we highlight that the requirement in equation (\ref{for:mdynmb}) implies that, when the PB stellar mass is doubled, also the dark-matter mass is doubled, for fixed truncation radius. Since at $\zi=20\kpc$ the effects of the dynamical friction are still non-negligible, we expect the PB to sink slowly towards the galaxy centre in a few $\Gyr$ (equation \ref{for:tdf}).

The PB ICs have been generated as described in Section~\ref{sec:PBICs}, considering its separate luminous and dark components. 
We verified that the PB is in equilibrium by evolving it for $10\Gyr$ in isolation. The only numerical effect is a slight decrease of the central density, which is negligible for the purposes of our investigation.

For clarity, we will refer to the two hydrodynamical $N$-body models as PBwithDM\_M0.5\_d20 and PBwithDM\_M1\_d20, where the terms 0.5 and 1 indicate the PB stellar mass in units of $10^8\Msun$, while d20 states that the PB centre of mass is located at an initial distance of $z=20\kpc$ (see Table~\ref{tab:sim2input}). As in the previous section, the ICs of NGC 5474 correspond to the equilibrium configuration of Appendix~\ref{sec:DHeq} after $\simeq0.98\Gyr$. Details on simulations and NGC 5474 parameters are listed in Tables~\ref{tab:param} and \ref{tab:siminput}. The simulations run for $7\Gyr$ and we use an adaptive timestep refinement with typical timestep values of $0.2\Myr$.

\begin{table}
    \centering
    \caption{Main input parameters of the set of simulations of Section~\ref{sec:sim2}. From top to bottom: name of the model (model's name); PB total stellar mass ($\Mb$); PB total dark-matter mass ($\MdmPB$); number of stellar particles used for the PB ($\Nb$); number of dark-matter particles used for the PB ($\Nbdm$); softening used for the PB stellar component particles ($\lPB$); softening used for the PB dark-matter component particles ($\lPBdm$). The softening is computed as in Section~\ref{sec:sim1pb} and all the simulations components (NGC 5474 dark halo, stellar disc, gas disc, PB stellar component and PB dark-matter halo) have different softenings. The PB stellar particles have $\mpart=1667\Msun$, while the dark-matter particles $\mpart=5000\Msun$. Details on how the PB dark-matter halo parameters have been fixed are in Section~\ref{sec:PB}, while details on the PB initial position and velocity are in Section~\ref{sec:sim2pb}. The ICs of NGC 5474 correspond to the configuration of Appendix \ref{sec:DHeq} taken after $0.98\Gyr$ (see also Tables~\ref{tab:param} and \ref{tab:siminput}).}\label{tab:sim2input}
    \begin{tabular}{lcc}
    \hline\hline
    Models' name    &   PBwithDM\_M0.5\_d20  &   PBwithDM\_M1\_d20  \\
    \hline\hline
    $\Mb$ [$10^8\Msun$] &   0.5     &  1        \\
    $\MdmPB$ [$10^9\Msun$]  &   0.98        &     1.95      \\
    $\Nb$               &   30000   &   60000   \\
    $\Nbdm$             &   195160  &   390320  \\    
    $\lPB$ [$\kpc$]     &   0.042   &   0.033   \\
    $\lPBdm$ [$\kpc$]   &   0.16    &   0.16    \\   
    \hline\hline
    \end{tabular}
\end{table}

\subsubsection{Results}
\label{sec:resPBDM}

\begin{figure*}
    \centering
    \includegraphics[width=1.\hsize]{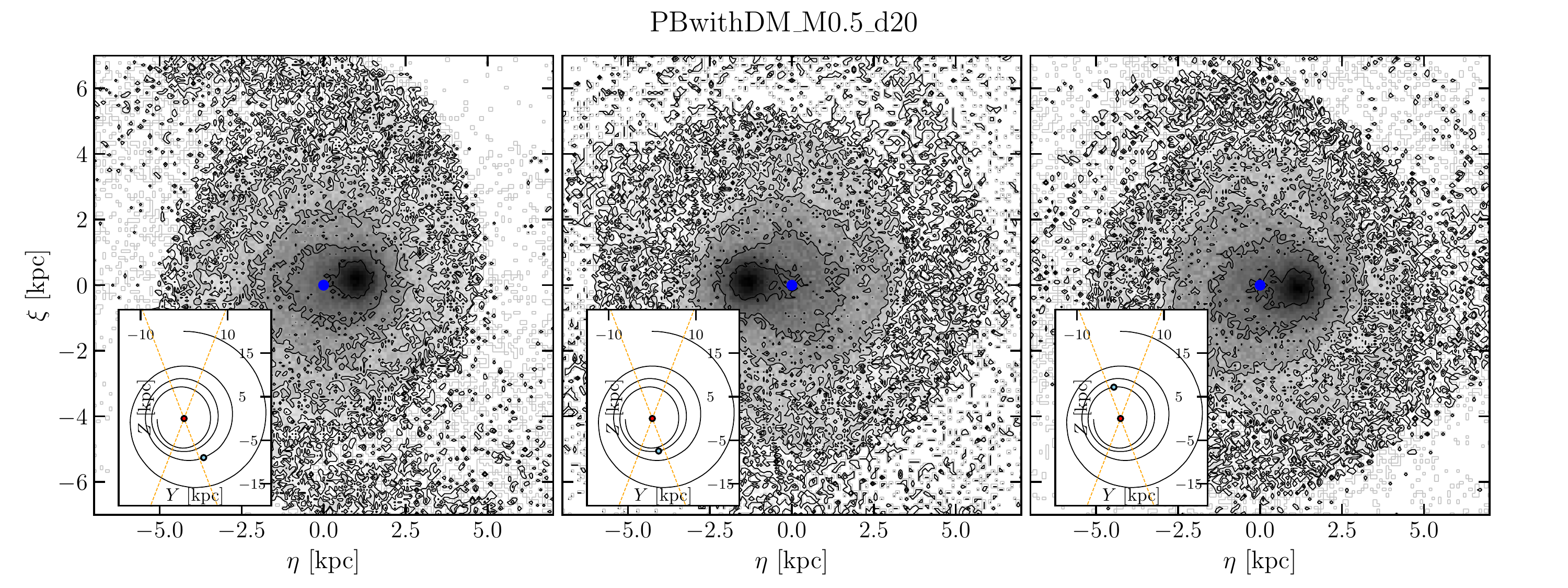}
    \caption{Total (stellar disc and PB) stellar surface density map taken at three snapshots of the model PBwithDM\_M0.5 d20 assuming an inclination $i = 21^{\circ}$. In each snapshot the PB produces, as a perspective effect, an off-set of 1 kpc on the discs’ plane. The centre of $(\eta,\xi)$-plane is the kinematic center of the stellar disc (blue dot), and the shades of greys extend for two order of magnitudes from the stellar density peak, corresponding, assuming a constant mass-to-light-ratio for both populations (stellar disc and PB) to a depth of five magnitudes. Indicating with $\Sigma_{\rm max}$ the densest peak, the contours are separated by $\Sigma_{\rm max}/2^n$, with $n=1,..., 6$. The three snapshots have been taken after the system has evolved for $\tint=3.43\Gyr$, $4.75\Gyr$ and $5.28\Gyr$, respectively from the left to the right panel. In the small insets we show the full PB orbit as in Fig.~\ref{fig:orbitDM}, marking with a blue dot the position of PB along its orbit that corresponds to its reference main panel, and with the dashed-orange curves the possible galaxy’s lines-of-sight.}
   \label{fig:PBoffset}
\end{figure*}

Figure~\ref{fig:orbitDM} shows the trajectories of the PBs of models PBwithDM\_M0.5\_d20 and PBwithDM\_M1\_d20, alongside their projected spatial distribution taken at two representative snapshots. The systems have been projected assuming as line of sight the galaxy's $x$-axis, i.e. the axis perpendicular to the plane of the orbit. The trajectories of the satellites differ the most in the number of windings around NGC 5474. This different behavior is completely driven by dynamical friction: since the PB of model PBwithDM\_M1\_d20 has twice the dynamical mass of model PBwithDM\_M0.5\_d20, its orbital time is approximately halved (equation \ref{for:tdf}). While the lighter PB (left panel) completes approximately four excursions above and below the equatorial plane in $\simeq6\Gyr$, the stellar component of the most massive PB (right panel) develops tidal tails after $\simeq3\Gyr$, when it has completed less than three windings. In Fig.~\ref{fig:orbitDM} we have marked the possible galaxy's inclinations ($i=21^{\circ}$, as in \citetalias{Rownd1994}) with two orange bands. The edges of such bands are such that, when crossed by the PB centre, they produce an off-set of $1\kpc$ as an effect of projection on the stellar disc of NGC 5474. In both panels the orbits are interrupted when the systems have reached $r\sim5-6\kpc$ from the centre since, at shorter distances, the stellar components of the PBs develop non-equilibrium features (tidal tails, elongated structures) apparently inconsistent with observations, so we do not include them in any following analysis.

The panels of Fig.~\ref{fig:PBoffset} show the total stellar (PB and disc) density map from three configurations from model PBwithDM\_M0.5\_d20, projected assuming an inclination $i=21^{\circ}$. The panels, from the left to the right, are projections taken after $\tint=3.43\Gyr$, $4.75\Gyr$ and $5.28\Gyr$, respectively after one and a half, two and a half and three full excursions above and below the equatorial plane. The gradient of contours is as in Fig.~\ref{fig:stellarzooms} and the small insets show the corresponding position of the PB along its orbit. When the PB crosses the galaxy's equatorial plane it causes, due to the gravitational perturbation, the development of a long-living and distinct spiral pattern in the stellar disc made up by two symmetric arms that dig the stellar disc up to $2\kpc$ and extend out to $5-6\kpc$. It is worth noticing that the spiral pattern does not form when the NGC 5474 galaxy model is evolved in isolation (see Appendix~\ref{sec:DHeq}, right panels of Fig.~\ref{fig:discmapeq}). 

\begin{figure}
    \centering
    \includegraphics[width=.9\hsize]{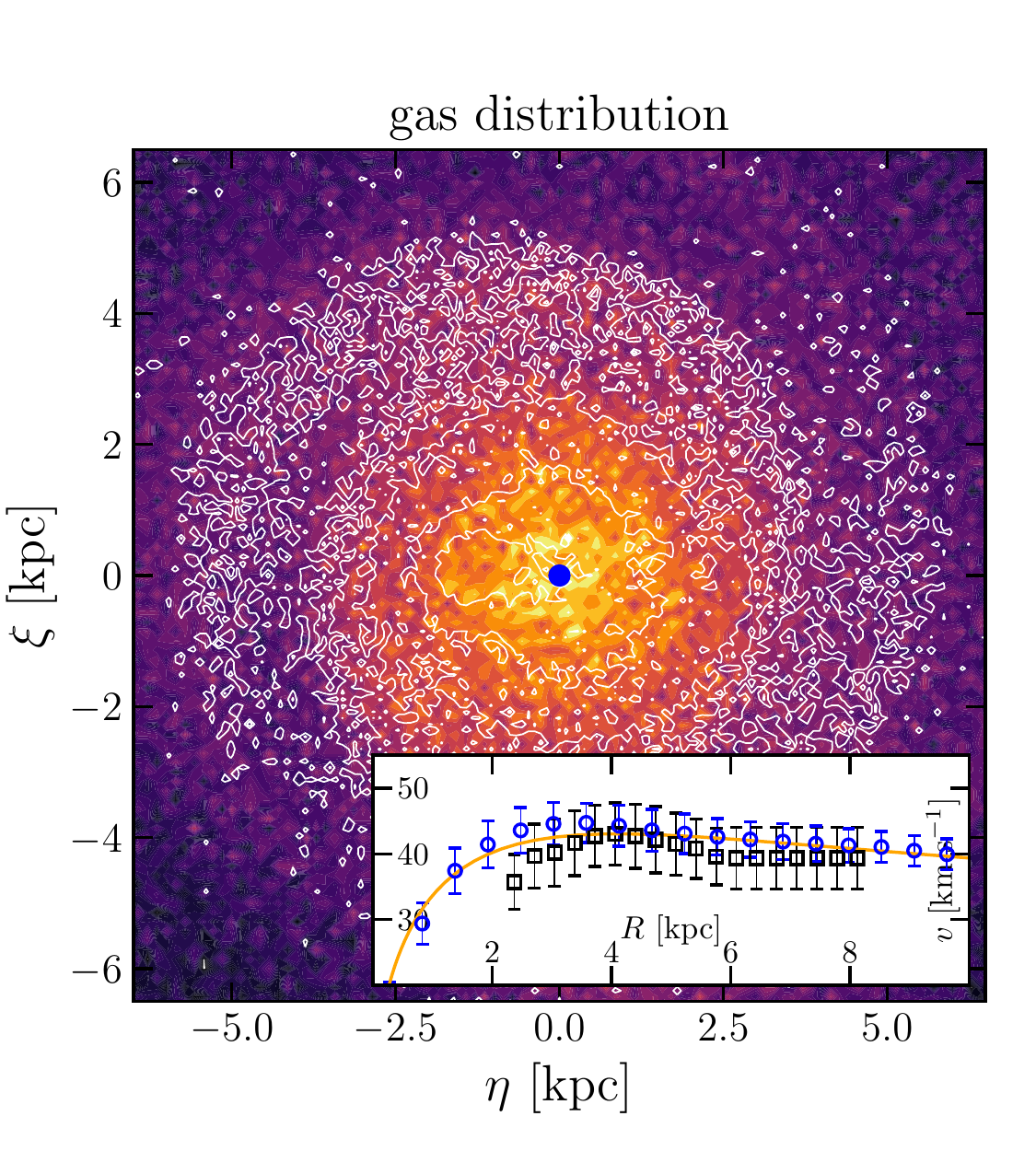}
    \caption{Same as the middle panel of Fig.~\ref{fig:PBoffset}, but showing the $\HI$ disc surface density map. The density decreases from yellow to blue. The white contours show the stellar spatial distribution of the configuration of the middle panel of Fig.~\ref{fig:PBoffset}, where, for clarity, we have removed the lowest isodensity contour. The blue dot shows the kinematic centre of the $\HI$ disc. Small inset: $\HI$ circular speed as a function of the galactocentric distance (blue points with error bars) from the same snapshot as in the main panel, compared to the galaxy's deprojected rotation velocity computed in Section~\ref{sec:hd} (black points with error bars). The orange curve shows the circular speed of the analytic model of NGC 5474.}
    \label{fig:PBoffsetgas}
\end{figure}

Similar spiral arms form also in the $\HI$ disc. Although much more structured and extended (we recall that $\hgas/\hstar\simeq4$), the pattern of the $\HI$ follows the one of the stellar component. As an example, Fig.~\ref{fig:PBoffsetgas} shows the $\HI$ projected density map corresponding to the middle panel of Fig.~\ref{fig:PBoffset}. As a reference, we have superimposed with white contours the projected stellar density map of the middle panel of Fig.~\ref{fig:PBoffset}. A spiral pattern forms also in model PBwithDM\_M1\_d20. In both models the arms develop after the satellite has crossed the discs plane at least at $\sim7\kpc$. This happens sooner in model PBwithDM\_M1\_d20, but it lasts for less due to the smaller dynamical friction time. At this distance, at least for model PBwithDM\_M0.5\_d20, the crossings of the equatorial plane do not perturb sensitively the kinematics of the $\HI$ disc, whose rotation curve (inset in Fig.~\ref{fig:PBoffsetgas}) still looks very similar to the the initial one and to the measured rotation curve of NGC 5474. When comparing with the observed morphology of NGC 5474, it is important to bear in mind that our models do not include star formation. For example, in the real galaxy, star forming and $\HII$ regions trace a different spiral pattern with respect to old-intermediate age stars. Our simulations can approximately trace the latter but not the former. 

\begin{figure}
    \centering
    \includegraphics[width=.9\hsize]{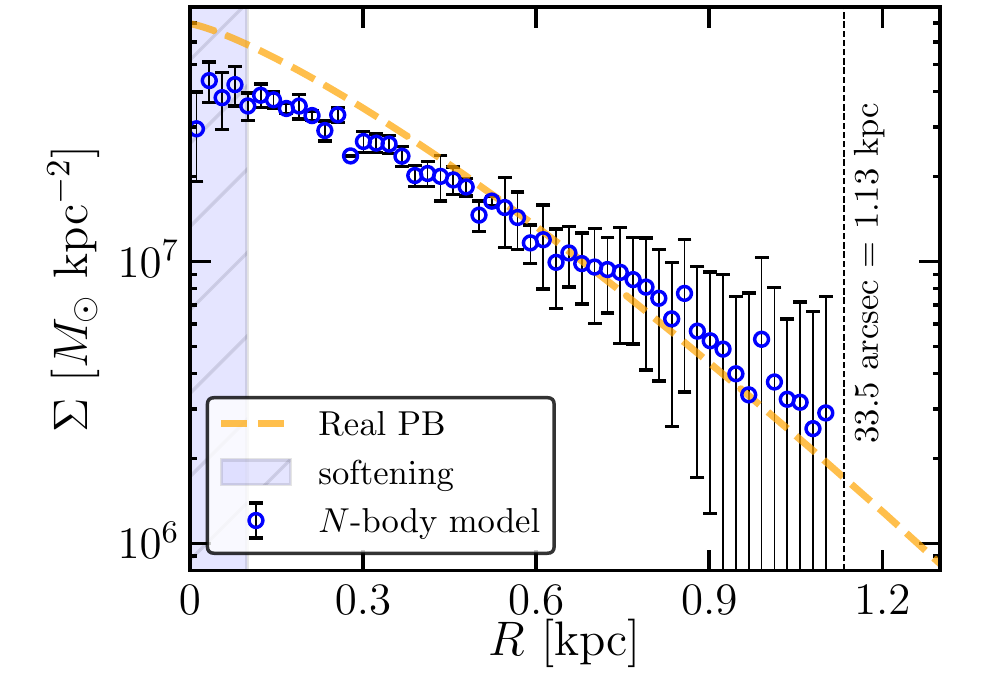}
    \caption{Projected density distribution of the PB as a function of the distance $R$ from the PB centre (blue circles with errorbars) superimposed to the analytic \Sersic model (yellow dashed line) from which the ICs have been sampled (i.e. with the same $m$ and $\Reff$ as the observed PB, \citetalias{Bellazzini2020}, but with a total initial mass $\Mb=0.5\times10^8\Msun$). The PB corresponds to the configuration taken as in the middle panel of Fig.~\ref{fig:PBoffset} and Fig.~\ref{fig:PBoffsetgas}.} .   
    \label{fig:mock}
\end{figure}

Figure~\ref{fig:mock} shows the PB stellar projected density profile obtained from the middle panel of Fig.~\ref{fig:PBoffset}. The profile has been computed as in the previous section by binning with 50 spherical annuli, equally spaced in $R$, out to $1.13\kpc$ (corresponding to $33.5\asec$, as in \citetalias{Bellazzini2020}). The background has been evaluated from a wider and distant annulus, and subtracted to the main profile. For comparison, the yellow curve shows the PB \Sersic model as in the analytic model from which the ICs have been sampled (corresponding to the \Sersic model of \citetalias{Bellazzini2020}, with a total stellar mass $\Mb=10^8\Msun$). The two profiles differ the most in the central parts, even though, as mentioned and discussed in the previous Section, we find the very same difference also when the system is evolved in isolation. 

\begin{figure}
    \centering
    \includegraphics[width=.8\hsize]{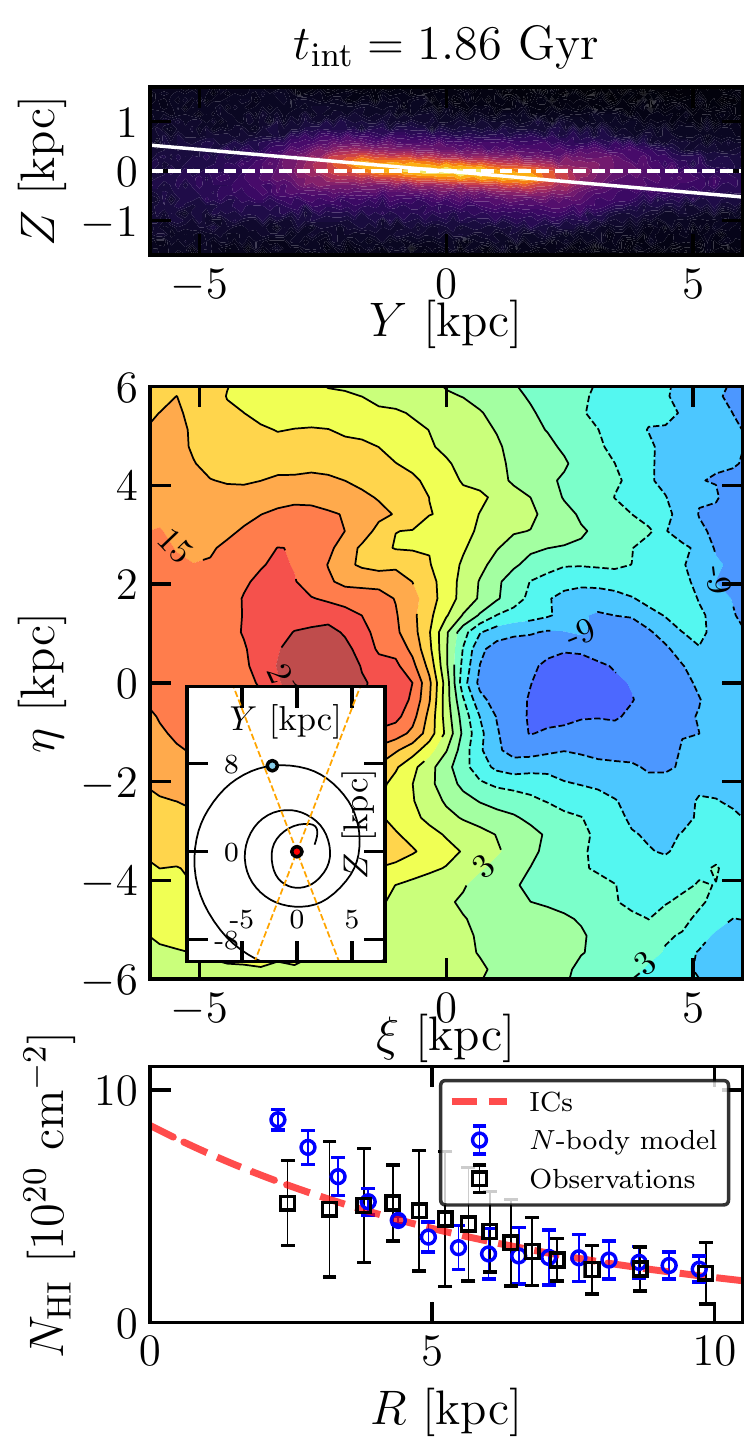}
    \caption{Top panel: $\HI$ projected density distribution when the galaxy is viewed edge-on. The configuration is taken from model PBwithDM\_M1\_d20 after $\tint=1.86\Gyr$ from the beginning of the simulation. The shades of colours, from yellow to black show regions of decreasing density, the dashed curve shows the $(X,Y)$-plane while the solid curve is $5^{\circ}$ inclined, as the warped $\HI$ disc. Middle panel: $\HI$ velocity field map from the same snapshot as in the top panel. The system is projected with an inclination of $i=21^{\circ}$ and the spatial and velocity resolutions are as in Fig.s~\ref{fig:stellarzooms} and ~\ref{fig:SW}. The small inset shows with a blue dot the corresponding position of the PB along its orbit (black circle) and with a red dot the galaxy's centre. The two orange lines mark the galaxy's possible inclinations. Bottom panel: projected HI column density profile computed from the same snapshot as in the top and middle panels (blue dot with errorbars) superimposed to the observed profile (black squares with errorbars), as derived in Section~\ref{sec:model}, and to the analytic models from which the ICs have been sampled (dashed red curve).}   
    \label{fig:warp}
\end{figure}

As a second, interesting feature, we find that the satellite mildly warps the $\HI$ disc when it crosses the galaxy's equatorial plane. The top panel of Fig.~\ref{fig:warp} shows the projected density distribution of the $\HI$ disc from model PBwithDM\_M1\_d20, when the galaxy is viewed edge-on, after the system has evolved for $\tint=1.86\Gyr$ and the PB has completed a full oscillation in the vertical direction, crossing the equatorial plane twice. We selected a snapshot in which the PB is also $1\kpc$ off-centred from the discs' centre, similarly to Fig.~\ref{fig:PBoffset} (see the small inset in the middle panel of Fig.~\ref{fig:warp}). The $\HI$ bends by approximately $5^{\circ}$ and the warp lives for another complete full vertical oscillation of the PB around NGC 5474. However, the $\HI$ velocity field strongly constrains the minimum distance that a system with such dynamical mass can reach. The middle panel shows the $\HI$ velocity field map from the same configuration of the top panel, but when the galaxy is seen inclined by $i=21^{\circ}$ (the resolution of the map is as in Fig.s~\ref{fig:stellarzooms} and ~\ref{fig:SW}). The iso-velocities contours are regular enough to be consistent with those of \citetalias{Rownd1994}, even though the approaching and receding arms are not symmetric over the full map: the approaching arm reaches a $21-24\kms$ amplitude at $2-3\kpc$, while it peaks at $12-15\kms$ at the same distance in the opposite direction. The maximum velocity difference between the approaching and receding arms reported by \citetalias{Rownd1994} is only $2-3\kms$. The configuration is the last still compatible and in good agreement with the observations: for longer times, further interactions between the two systems erase any sign of regularity and differential rotation from the $\HI$ velocity field. The PB of model PBwithDM\_M0.5\_d20 does not warp or distort the $\HI$ disc when it crosses through it, keeping the $\HI$ velocity field regular, thanks to its lower dynamical mass and to the fact that most of the crossings happen for distances larger than $7\kpc$. We recall that in model PBwithDM\_M1\_d20, the PB dynamical mass is considerably high, more or less comparable to the one of NGC 5474. So, it sounds plausible that a system with a lower mass (in between the two models) could, at the same time, warp the $\HI$ without significantly distorting the $\HI$ velocity field map.

The bottom panel of Fig.~\ref{fig:warp} shows the $\HI$ column density distribution corresponding to the middle panel, compared with the $\HI$ column density we derived in Section~\ref{sec:model} from observations. The density profile of the $\HI$ changes and bends at $\sim5-6\kpc$, which corresponds approximately to the distance of the latest crossing of the PB, but the overall shape is consistent with the observed one, apart from the centre, where the two profiles differed the most already at the beginning of the simulation (see Appendix~\ref{sec:DHeq}, Fig.~\ref{fig:discseq}).


\section{Conclusions}
\label{sec:concl}

As a member of the M 101 Group, and appearing in projection so close to its giant central galaxy, a tumultuous past has always been invoked as the main driver of all of the peculiarities of NGC 5474. However, while the hypotheses of a gravitational interaction with M 101 may explain, for instance, the warped $\HI$ distribution \citepalias{Rownd1994}, it does not look like an explanation for its off-centred bulge. Off-set bars are observed in Magellanic spirals \citep{Odewahn1989}, even though the mechanism that can induce the misplacement is still unknown. NGC 5474 is, however, not a Magellanic spiral: i) it does not possess a bar, but rather a very round and regular stellar component; ii) it has two spiral arms and not one; iii) off-centred bars in Magellanic spirals are observed mostly in binary systems. To complicate things, the only available $\HI$ observations of NGC 5474 date back to the early 90s, and only trace the large scale structure of the galaxy, while the more recent $\Halpha$ observations, tracing the inner kinematics, seem to be hardly reconcilable with the $\HI$ data \citepalias{Epinat2008,Bellazzini2020}, though the galaxy's low inclination does not allow to draw robust conclusions on the disc's kinematics.

Following the work of \citetalias{Bellazzini2020}, who renewed the interest in this galaxy accomplishing a detailed study of its stellar populations, we have produced state-of-the-art hydrodynamical $N$-body models of NGC 5474, aimed to investigate the nature of the galaxy's central and compact stellar component, usually interpreted as an off-set bulge. Using analytic models we have argued that, if the PB really lies within the galaxy's disc plane, it is implausible that its dynamical mass is more than $10^8\Msun$, because such a system would: i) shift the entire galaxy's gravitational potential minimum, making the kinematic centre of the discs coincide with the centre of the bulge; ii) induce strong distortions in the $\HI$ velocity field map, inconsistent with observations. 

Through hydrodynamical $N$-body simulations, we tried to reproduce configurations where the PB appears off-centred as a result of orbital decay due to dynamical friction, when it moves within the galaxy's discs plane. We explored PBs of different masses and sizes but, in none of the considered scenarios we were able to reproduce the observations: a system with a large initial size ($\Reff\ge320\kpc$), while evolving into the strong tidal force field of NGC 5474, develops massive tidal tails and gets flattened and elongated (in some cases destroyed) after less than $0.8\Gyr$. The very short time needed to show non-equilibrium features makes very unlikely that a PB with these characteristics can either come from the galaxy's outer regions or be an off-centred pseudo-bulge. A compact system reaches the required $1\kpc$ distance from the centre but, on the basis of structural analysis, it remains either too compact or gets too flattened to look similar to the observed one.

Through $N$-body simulations, \cite{Levine1998} showed that an off-set between a stellar disc and the gravitational potential minimum of its host galaxy can stand for sufficiently long time if the stellar disc spins in a sense retrograde to its orbit about the halo centre. According to the authors, we should then interpret the discs as off-centred with respect to the bulge, and not the other way around. We believe that this is not the case of NGC 5474 in which the off-set stands in the gas kinematics as well (\citealt{Levine1998} considered collisionless simulations with no gas). If we imagine the gas to behave similarly to the stellar counterpart, according to \cite{Levine1998}, the $\HI$ velocity field should be clearly and strongly asymmetric, which is not the case for NGC 5474.

As different authors proposed (\citetalias{Rownd1994}; \citealt{Mihos2013}, \citetalias{Bellazzini2020}), we have explored the hypothesis that the off-set is produced by projection effects, once the PB is orbiting around NGC 5474. Due to the structural homology between the PB and a dE galaxy, we have coated it with dark-matter halo. We have shown reference cases of polar orbits where, in projection, the PB looks off-centred of $1\kpc$, just as observed. We exploit the gravitational interaction between the satellite PB and NGC 5474 to show that it may: i) explain the formation of the galaxy loose spiral pattern, formed by two symmetric arms, together with a very similar structure in the $\HI$ distribution; ii) partially account for the formation of the warped $\HI$ disc, at least in cases of sufficiently massive PB. 

Of course, the large parameter space, the lack of tight observational constraints and the degeneracy induced by projection would allow hundreds of orbits to reproduce the observed, present-day configuration. As such, we do not expect to have solved all the mysteries behind NGC 5474, but rather to have shown in a quantitative manner that its PB is probably not the bulge or the pseudo-bulge of NGC 5474, and we have also presented a possible, intriguing alternative scenario where the PB is a satellite galaxy of NGC 5474, moving on a polar orbit, that has the advantage of explaining some of the other peculiarities of NGC 5474.

While our study is not sufficient to ascertain the real nature of the PB, it provides for the first time a sound way out to the main puzzle of the structure of NGC 5474: the odd off-centred \lapex bulge \rapex is likely not a bulge at all, but a satellite dwarf galaxy projected near the center of a M33-like bulge-less spiral \citep{Boker2002,Das2012,Grossi2018}.

\section*{Acknowledgments}
We thank the anonymous referee for his/her comments and suggestions that considerably improved the quality of this work. We acknowledge the use of computational resources from the parallel computing cluster of the Open Physics Hub (\url{https://site.unibo.it/openphysicshub/en}) at the Physics and Astronomy Department in Bologna. We acknowledge funding from the INAF Main Stream program SSH 1.05.01.86.28. FM is supported by the Program \lapex Rita Levi Montalcini\rapex of the Italian MIUR. We thank F. Fraternali and G. Iorio for very helpful discussions. RP acknowledges G. Sabatini for useful suggestions and comments.

\section*{Data availability}
The rotation curves and the $\HI$ column density map of NGC 5474 are available at https://dx.doi.org/10.1086/117185. The rotation curve and the $\HI$ density distribution rederived in this article will be shared on request to the corresponding author.

\bibliography{paper}
\bibliographystyle{mnras}

\appendix
\section{The initial conditions of NGC 5474}
\label{sec:DHeq}

\begin{figure}
 \centering
 \includegraphics[width=1\hsize]{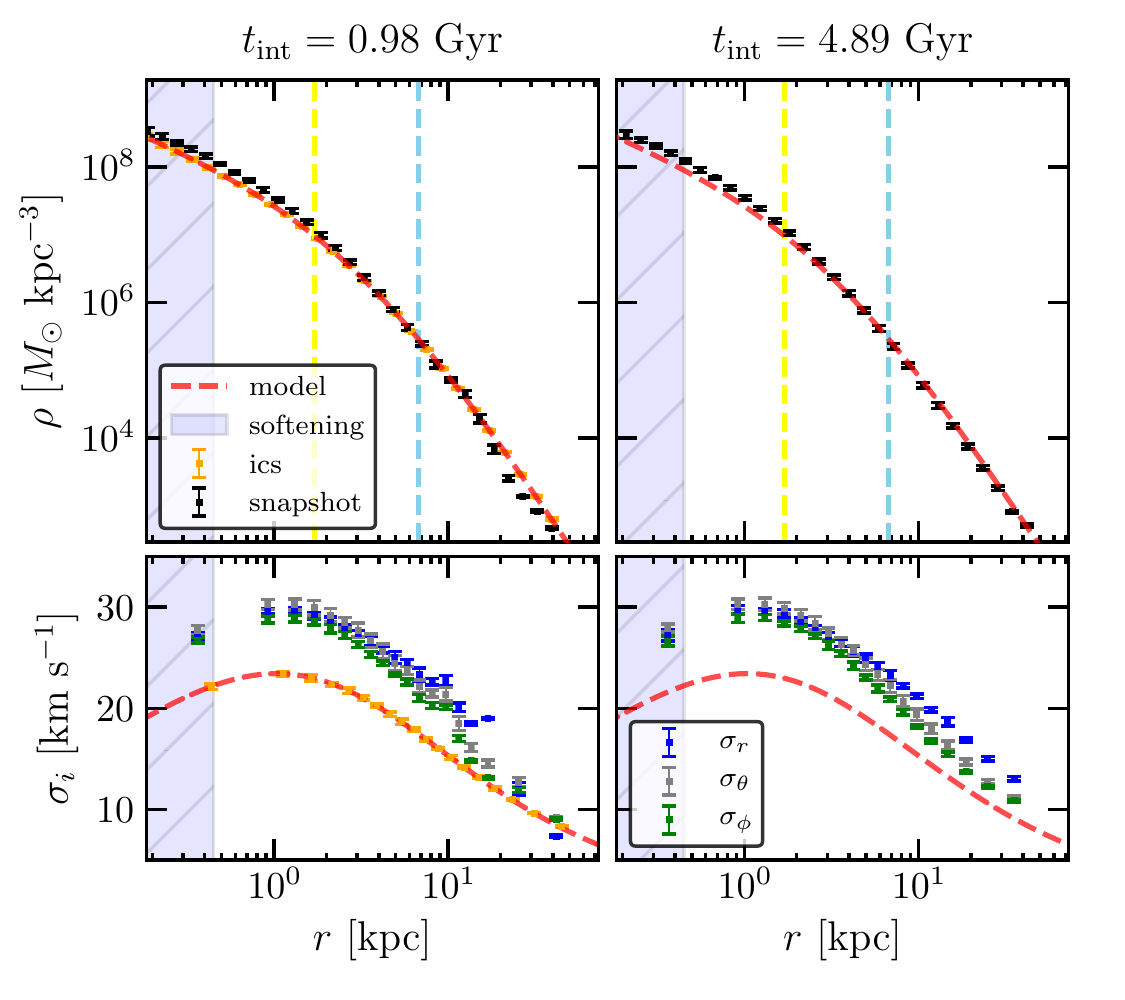}
 \caption{Top panels: density distribution of the halo (black points with errorbars) from two reference snapshots of the NGC 5474 simulation of Appendix~\ref{sec:DHeq}, corresponding to $\tint=0.96\Gyr$ (left column of panels). and $\tint=4.89\Gyr$ (right column of panels). Bottom panels: velocity dispersion tensor elements $\sigmai$, with $i=r$ (blue points with errobars), $\theta$ (grey points with errobars), and $\phi$ (green points with errobars). The left-hand panels also show the corresponding quantities as computed from the ICs (orange points with error bars), while the analytic density distribution (top panels) and the velocity dispersion of the isotropic analytic model (bottom panels) from which the halo ICs have been sampled is shown with a dashed-red curve. The radial bins in the velocity dispersion profiles have been computed so that they contain the same number of particles $\Nh/20$. The light blue bands in all panels extend out to $3\ldm$, with $\ldm$ the adopted softening (see Table~\ref{tab:siminput}). The yellow and blue dashed vertical lines show, respectively, the region marked by the stellar ($\hstar$) and $\HI$ ($\hgas$) discs scale lengths. 
 }\label{fig:haloeq}
\end{figure}

\begin{figure}
 \centering
 \includegraphics[width=.9\hsize]{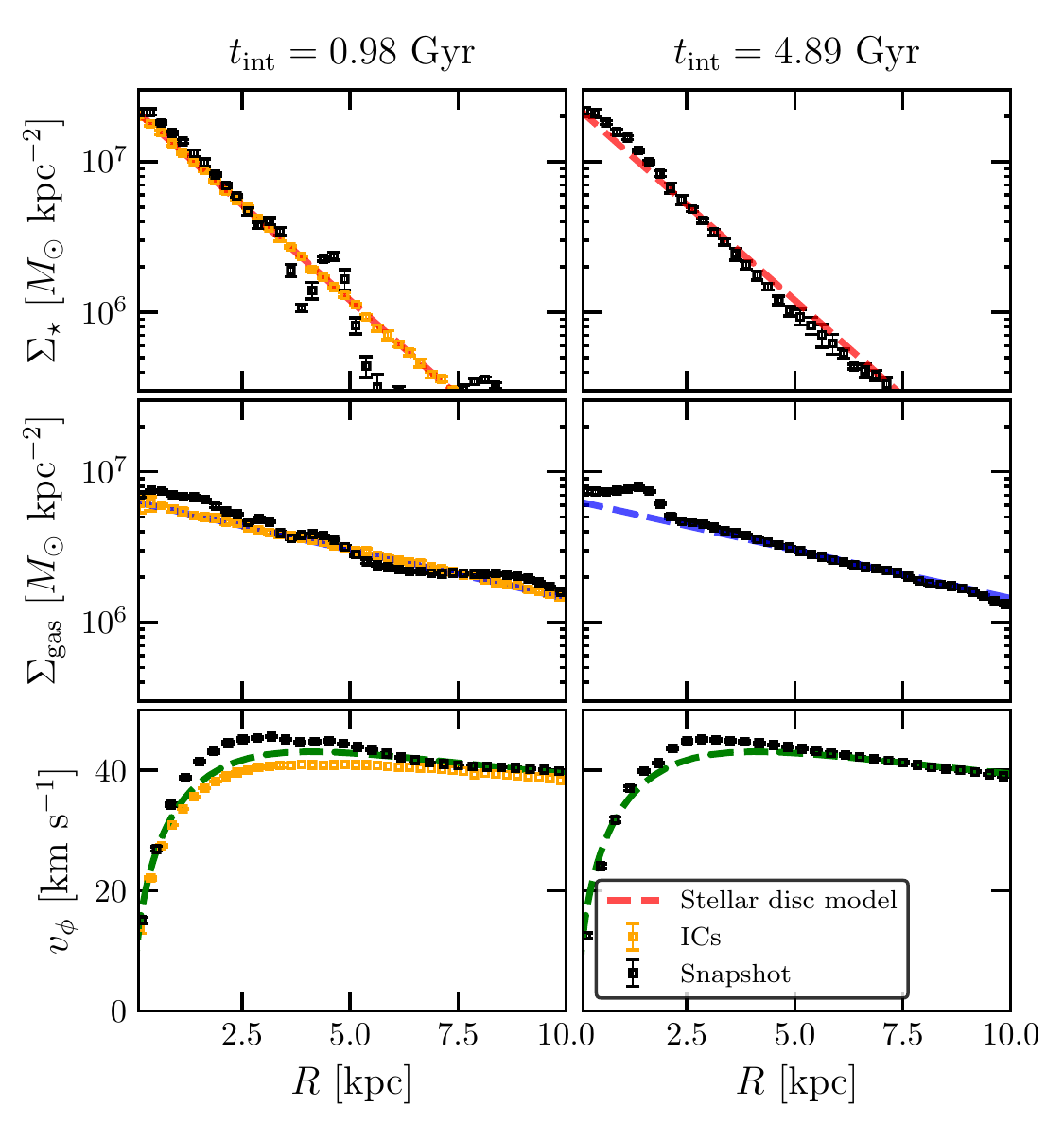}
 \caption{Main structural and kinematic properties of the stellar and $\HI$ discs from two reference snapshots of the NGC 5474 simulation of Appendix~\ref{sec:DHeq}, corresponding to $\tint=0.98\Gyr$ (left column of panels) and $\tint=4.89\Gyr$ (right column of panels). Top panels: stellar disc surface density profile (black points with error bars); middle panels: $\HI$ projected density distribution (black points with error bars); bottom panels: $\HI$ azimuthal velocity curve (black points with error bars). The system has been projected assuming the symmetry axis as line of sight. In the left column we also show the corresponding quantities as derived from the ICs (orange points with error bars). The dashed red and blue curves show, respectively, the stellar and $\HI$ surface density as from the analytic model from which the ICs have been sampled, while the dashed green curve in the bottom panels show the analytic model circular speed. 
 }\label{fig:discseq}
\end{figure}

\begin{figure*}
 \centering
 \includegraphics[width=.48\hsize]{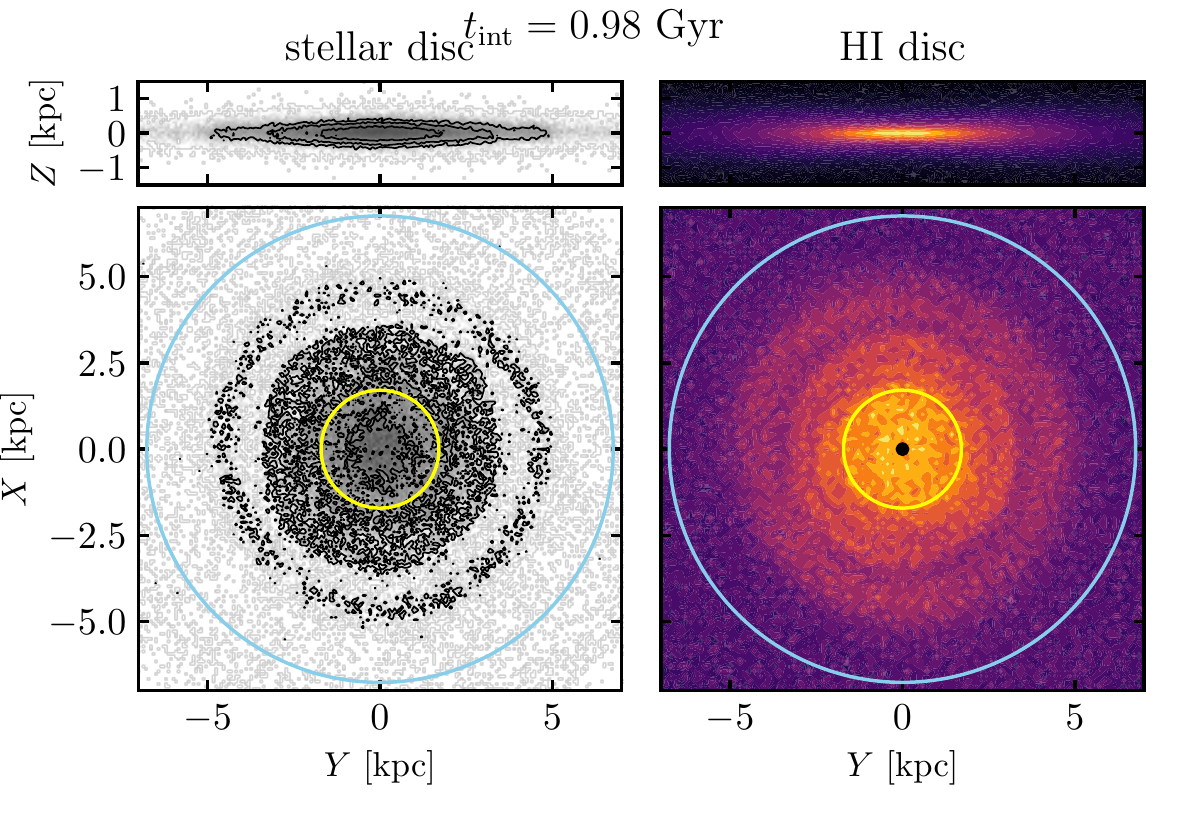}
 \includegraphics[width=.48\hsize]{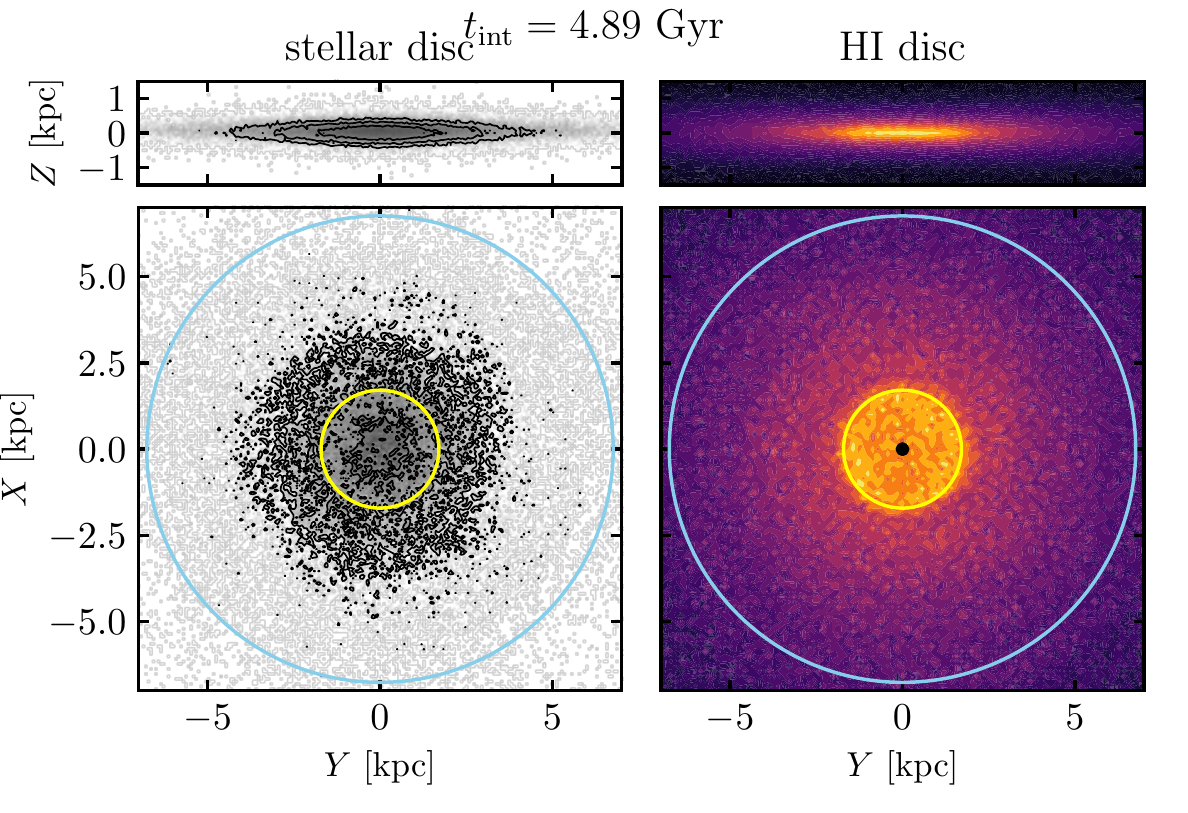}
 \caption{Stellar and $\HI$ discs main structural and kinematic properties. The two left-hand columns of panels show the face-on (bottom panels) and edge-on (top panels) stellar disc and $\HI$ surface density maps, projected along the galaxy's symmetry axis. The configuration is taken after the system has evolved for $\tint=0.98\Gyr$. The two right-hand columns of panels show the same as in the left panels but after the system has evolved for $\tint=4.89\Gyr$. The yellow and blue circles show, respectively, the region marked by the stellar and $\HI$ discs scale lengths. The snapshot are taken from the simulation of NGC 5474 of Appendix~\ref{sec:DHeq}. 
 }\label{fig:discmapeq}
\end{figure*}

\begin{table}
    \centering
    \begin{tabular}{rlcl}
    \hline\hline
           parameter        &   DM halo       &   stellar disc  &   $\HI$ disc              \\
    \hline\hline
    ($\Mdm,\Mstar,\Mgas$) =   &    $(3.77$,     &   $0.41,$       &   $1.82)\times10^9\Msun$  \\
    ($\Nh,\Nstar,\Ngas$)  =  &    $(708750$,   &   $82200,$     &   $364000)$               \\
    ($\ldm,\lstar,\lgas$) =  &    $(0.15$,     &   $0.06,$       &   $0.075)\kpc$            \\
    \hline\hline
    \end{tabular}
    \caption{Main input parameters of the hydrodynamical $N$-body simulation of NGC 5474 in isolation. $\Mdm$, $\Mstar$, $\Mgas$: respectively total dark-matter, stellar and $\HI$ masses. $\Nh$, $\Nstar$, $\Ngas$: number of particles used for the dark-matter halo, the stellar disc and the $\HI$ disc. $\ldm$, $\lstar$, $\lgas$: respectively the softening used in all the simulations for the halo, the stellar and the $\HI$ discs. The softenings are computed requiring that the maximum force between the component's particles should not be larger than the component's mean-field strength \citep{Power2003,Dehnen2011}. All particles have the same mass $\mpart=5000\Msun$.}\label{tab:siminput}
\end{table}

The ICs of the NGC 5474-like galaxy model, as sampled in Section~\ref{sec:comp}, are evolved in isolation for $\tmax=10\Gyr$ to check how the components respond to the presence of each other, and to let the system, which is built in quasi-equilibrium, shift towards equilibrium. We use an adaptive timestep refinement, with typical timesteps of $0.3\Myr$. Table~\ref{tab:siminput} lists some of the main input parameters used to generate the ICs of the galaxy model of NGC 5474 and run the simulation (masses, number of particles and softenings).


In Fig.~\ref{fig:haloeq} we show the dark-matter density (top panels) and velocity dispersion tensor elements $\sigmar$, $\sigmatheta$ and $\sigmaphi$ (bottom panels) and compare them with the corresponding quantities from the dark-matter analytic model. The profiles are computed after the galaxy has evolved for $\tint=0.98\Gyr$ (left column) and $\tint=4.89\Gyr$ (right column). The halo density increases by approximately a factor of two while the velocity dispersion components $\sigma_i$ (with $i=r,\theta,\phi$) increase by $6-7\kms$, in a region confined mostly within $\hstar$, where the discs contribute the most. The increase in the velocity dispersions is of a factor 1.3, and the overall profiles are approximately the same one would get with an isotropic Hernquist model with $\Mtot=\Mdm+\Mstar+\Mgas$. For a spherical system, the parameter 
\begin{equation}\label{for:beta}
    \beta = 1 - \frac{\sigmatheta^2+\sigmaphi^2}{2\sigmar^2} = 1-\frac{\sigmatheta^2}{\sigmar^2}
\end{equation}
measures the models' anisotropy distribution\footnote{We recall that spherical symmetry implies $\sigmatheta=\sigmaphi$.}. An isotropic model corresponds to $\beta=0$, while $\beta<0$ and $0<\beta\le1$ indicate, respectively, tangential and radial velocity distributions. We also note that the halo changes its velocity distribution from isotropic ($\beta=0$) to radially biased ($\beta>0$) due to the weak collapse induced by the deeper potential well.

In Fig.~\ref{fig:discseq} we show some of the main structural and kinematic properties of the stellar and $\HI$ discs corresponding to the same configurations as in Fig.~\ref{fig:haloeq}. The top rows show the stellar projected density distribution, the middle row the $\HI$ projected density distribution, while the $\HI$ streaming velocity profile is shown in the bottom panels. As for the halo, the discs adjust in the very beginning of the simulation: they respond to the contraction of the halo increasing their central densities, in a region extending out $R\simeq\hstar$, until a new equilibrium configuration is reached. As a result of the deeper potential well caused by the collapse of the halo in the galaxy's central parts, mostly the stellar disc develops radial density waves that are still in place after $1\Gyr$, although in a low density region. Figure~\ref{fig:discmapeq} shows the discs density maps, face-on and edge-on, computed from the same configurations as in Figs~\ref{fig:haloeq} and \ref{fig:discseq}. Both discs are generated with non negligible thickness. We recall that the stellar disc vertical profile follows from equation (\ref{for:fullstar}), with $\zstar=0.15\hstar$, while the thickness of the gaseous disc is determined by means of the vertical hydrostatic equilibrium (\ref{for:idroeq}). The radial density wave is clearly visible as the outer ring in the stellar surface density map of the left panel of Fig.~\ref{fig:discmapeq}. Also, we notice that the stellar disc does not develop any spiral structure along the whole simulation.

\label{lastpage}

\end{document}